\documentclass[a4paper]{article}
\usepackage[english]{babel}
\usepackage{authblk}
\usepackage{natbib}
\usepackage{graphicx}
\usepackage{bbold}
\usepackage{a4wide}
\usepackage{verbatim}
\usepackage{color}

\begin{document}
\title{Revisiting enumerative two-part crude MDL \\
for Bernoulli and multinomial distributions \\
(Extended version)}
\author{Marc Boull\'e, Fabrice Cl\'erot, Carine Hue}
\affil{Orange Labs - 22300 Lannion - France}
\maketitle

\abstract{
We leverage the Minimum Description Length (MDL) principle as a model selection technique for Bernoulli distributions and compare several types of MDL codes.
We first present a simplistic crude two-part MDL code and a Normalized Maximum Likelihood (NML) code.
We then focus on the enumerative two-part crude MDL code, suggest a Bayesian interpretation for finite size data samples, and exhibit a strong connection with the NML approach. We obtain surprising impacts on the estimation of the model complexity together with superior compression performance.
This is then generalized to the case of the multinomial distributions.
Both the theoretical analysis and the experimental comparisons suggest that one might use the enumerative code rather than NML in practice, for Bernoulli and multinomial distributions.
}

\section{Introduction}
\label{secIntroduction}

Model selection is a key problem in statistics and data mining, and the MDL approaches \citep{Rissanen78} to model selection have been extensively studied in the literature \citep{Grunwald07}, with successful applications in many practical problems.
Simple models such as Bernoulli or mainly multinomial distributions are important because they are easier to analyze theoretically and useful in many applications.
For example, the multinomial distribution has been used as a building block in more complex models, such as naive Bayes classifiers \citep{MononenEtAl07}, Bayesian networks \citep{RoosEtAl08}, decision trees \citep{VoisineEtAlAKDM09} or coclustering models \citep{BoulleHOPR10,GuigouresEtAlECML15}. These models involve up to thousands of multinomials blocks, some of them with potentially very large numbers of occurrences and outcomes. For example, the text $\times$ word coclustering of the 20-newsgroup dataset described in \citep{BoulleHOPR10} exploits a main multinomial block with around two millions words (occurrences) distributed on 200,000 coclusters (outcomes). In \citep{GuigouresEtAlECML15}, half a billion call detail records (occurrences) are distributed on one million coclusters (outcomes).
These various and numerous applications critically rely on the use of effective and efficient MDL code lengths to get a robust and accurate summary of the data.

The MDL approaches come with several flavors, ranging from theoretical but not computable to practical but sub-optimal.
Ideal MDL \citep{VitanyiEtAl00} relies on the Kolmogorov complexity, that is the ability of compressing data using a computer program. However, it suffers from large constants depending on the description method used and cannot be computed, not even approximated in the case of two-part codes \citep{AdriaansEtAl07}.
Practical MDL leverages description methods that are less expressive than general-purpose computer languages.
It has been employed to retrieve the best model given the data in case of families of parametrized statistical distributions.
Crude MDL is a basic MDL approach with appealing simplicity.
In two-part crude MDL, you just have to encode the model parameters and the data given the parameter, with a focus on the code length only. However, crude MDL suffers from arbitrary coding choices.
Modern MDL relies on universal coding resulting in Refined MDL \citep{Grunwald07}, with much stronger foundations and interesting theoretical properties.
In this paper, we investigate the enumerative two-part crude MDL code for the Bernoulli and multinomial models, exhibit a strong connection with the NML approach, with surprising impacts on the estimation of the model complexity and superior compression performance.

The rest of the paper is organized as follows.
For self-containment reasons, Section~\ref{secBernoulli} presents standard codes for the Bernoulli distribution: one simplistic two-part crude MDL code as well as a refined MDL code based on the NML approach.
Section~\ref{secRevisitedEnumerative} describes a particular two-part crude MDL code based on enumerations and establishes the connection of its parameter coding length with its NML parametric complexity.
Section~\ref{secComparison} proceeds with a deep comparison between this enumerative MDL code and the NML code presented in Section~\ref{secBernoulli}.
Section~\ref{secMultinomial} suggests an extension of the enumerative two-part crude MDL code to multinomial distributions and Section~\ref{secComparisonM} compares this code with the alternative NML code.
Finally, Section~\ref{secConclusion} summarizes this paper.

\section{Standard MDL codes for Bernoulli strings}
\label{secBernoulli}

We briefly present one simplistic example of two-part crude MDL code for encoding binary strings using the Bernoulli model, as well as a modern MDL code based on NML. This has been presented many times in the literature, e.g. \citep{HansenEtAl01b, Grunwald07, RooijEtAl09}.

Let us consider the Bernoulli model with $\theta \in [0, 1]$ in the case of binary sequences $x^n \in X^n$ of size $n$. Let $k(x^n)$ be the number of ones in $x^n$.

\subsection{Simplistic two-part crude MDL approach}
\label{secSimplistic}

Using a two-part version of the MDL principle \citep{Grunwald07}, we select the best hypothesis $H$ that explains the data $D$ by minimizing the sum $L(H) + L(D|H)$, where $L(H)$ is the coding length of the hypothesis and $L(D|H)$ is the coding length of the data encoded with the help of the hypothesis.

In the case of the Bernoulli model, we have to encode the parameter $\theta$ and the data $x^n$ given $\theta$.
The number of ones in the binary string $x^n$ is between 0 and $n$.
The $\theta$ parameter can thus be chosen among $(n+1)$ values $\theta=\frac {0} {n}, \frac {1} {n}, \frac {2} {n}, \ldots, \frac {n} {n}$, and be encoded using $L(\theta) = \log (n+1)$ bits.

For $\theta \in \{0, 1 \}$, the string $x^n$ is degenerated with only zeros or ones, and its coding length given $\theta$ is $L(x^n|\theta) = 0$.

For $\theta = \frac {k} {n},\; 0 < k < n$, every symbol of the string $x^n$ can be encoded using $-\log \frac {k} {n}$ bit for a one and $-\log \frac {n-k} {n}$ bit for a zero, leading to $L(x^n|\theta) = -k \ln \frac {k} {n} - (n-k) \ln \frac {n-k} {n}$.

This gives a total code length of

\begin{equation}
L(\theta = \frac {k} {n}, x^n) = \log(n+1) + \left(-k \log \frac {k} {n} - (n-k) \log \frac {n-k} {n}\right).
\end{equation}

Equivalently, the likelihood of the whole string $x^n$ can be estimated as $P(x^n|\theta = \frac {k} {n}) = (\frac {k} {n})^k (\frac {n-k} {n})^{n-k}$, with $L(x^n|\theta) = - \log P(x^n|\theta = \frac {k} {n})$. 

Using the Shannon entropy $H(\frac {k} {n}) = -\frac {k} {n} \log (\frac {k} {n}) - \frac {n-k} {n} \log (\frac {n-k} {n})$, we also have  $L(x^n|\theta) = n H(\theta)$.

\subsection{Standard NML Approach}
\label{secNML}

The simplistic two-part MDL code defined previously suffers from some arbitrary choices and may be suboptimal at best, with arbitrary bad behavior for small sample sizes \citep{Grunwald07}.

In the case of the Bernoulli model, this is pointed out in \citep{RooijEtAl09},
\begin{quotation}
``Example 5. $\ldots$
A uniform code uses $L(\theta) = \log(n + 1)$ bits to identify an element of this set. Therefore the resulting regret is always exactly $\log(n + 1)$. By using a slightly more clever discretisation we can bring this regret down to about $\frac {1} {2} \log n + O(1)$, which we mentioned is usually achievable for uncountable single parameter models.''
\end{quotation}

Using universal coding, a much more grounded approach is proposed to better evaluate the model complexity, based on the Shtarkov NML code, which provides strong theoretical guarantees \citep{Rissanen00}.

It exploits the following NML distribution $\overline{P}_{nml}^{(n)}$ on $X^n$:

\begin{equation}
\label{eqnNML}
\overline{P}_{nml}^{(n)}(x^n) = \frac { P_{\widehat{\theta}(x^n)}(x^n)} 
{\sum_{y^n \in X^n} {P_{\widehat{\theta}(y^n)}(y^n)}}
\end{equation}

where $\widehat{\theta}(x^n)$ is the model parameter that maximizes the likelihood of $x^n$.

The log of the denominator stands for the \textit{parametric complexity} $COMP^{(n)}(\theta)$ of the model whereas the negative log of the numerator is the \textit{stochastic complexity} of the data given the model.
The sum of both terms provides the NML code. It is noteworthy that the NML code is a one-part rather than two-part code: data is encoded with the help of all the model hypotheses rather than the best hypothesis.

\medskip

In the case of the Bernoulli model, $\widehat{\theta}(x^n) = k(x^n)/n$.
We have

\begin{eqnarray}
COMP^{(n)}(\theta) &=& \log \sum_{y^n \in X^n} {P_{\widehat{\theta}(y^n)}(y^n)}, \\
  &=& \log \sum_{k=0}^n {{{n}\choose{k}} \left(\frac {k}{n}\right)^k \left(\frac {n-k}{n}\right)^{n-k}}.
\end{eqnarray}

Using the Stirling' formula together with the Fisher information provides the following accurate approximation \citep{Rissanen96}:
\begin{eqnarray}
COMP^{(n)}(\theta) &=& \frac{1}{2} \log \frac{n}{2 \pi} + \int_{\theta} {\sqrt{ det I(\theta)}} +o(1),\\
 &=&  \frac{1}{2} \log \frac{n \pi}{2} + o(1).
\end{eqnarray}

Remarkably, this is in line with the classical BIC regularization term $\frac{1}{2}\log n$.

\section{Revisiting enumerative two-part crude MDL}
\label{secRevisitedEnumerative}

We present the enumerative two-part crude MDL code for Bernoulli distributions, suggest a finite data sample Bayesian interpretation and show a connection with the NML approach.

\subsection{Enumerative two-part crude MDL}
\label{secEnumerativeB}

We present an alternative type of two-part crude MDL code for Bernoulli distributions. It has already been proposed in the past literature, under the names of \emph{index} or \emph{enumerative} code (see for example \citet{Grunwald07} Example~10.1 \emph{Coding by Giving an Index}). 

First, like in Section~\ref{secSimplistic}, we enumerate all possible $\theta = \frac{i}{n}$ parameter values given the sample size $n$. We then use $\log (n+1)$ bits to encode $\theta$.
Second, given $\widehat{\theta}(x^n) = \frac{k(x^n)}{n}$, we enumerate all the ${{n}\choose{k}}$ binary sequences with $k$ ones and encode the data $x^n$ using $\log {{n}\choose{k}}$ bits.
This gives a total code length of

\begin{equation}
L(\widehat{\theta}(x^n), x^n) = \log(n+1) + \log \frac{n!} {k! (n-k)!},
\end{equation}

Interestingly, this crude MDL approach results in the same code length as that obtained in \citep{HansenEtAl01b} using \emph{Predictive Coding} or \emph{Mixture Coding} with a uniform prior.

For the case of the Bayes mixture model with uniform prior $w(\theta) = 1,\; \theta \in [0,1]$,
we have 
\begin{eqnarray}
P_{Bayes}(x^n) &=& \int_0^1{w(\theta) P_{\theta}(x^n)d\theta}, \\
 &=& \int_0^1{\theta^{k} (1-\theta)^{n-k} d\theta},\\
 &=& \frac{1}{n+1} \frac{k! (n-k)!} {n!} .
\end{eqnarray}
The negative log of $P_{Bayes}(x^n)$ actually corresponds to the code length of the enumerative code.

This code has also been studied by \citet{Grunwald07} (Chapter~10, Section~10.2) under the name \emph{Conditional Two-Part Universal Code}, which suggests that at least for the Bernoulli model, this code is strictly preferable to the ordinary two-part code.

\subsection{Bayesian interpretation}
\label{secEnumBayesian}

Let $\mathcal{M} = \{ P_\theta \, | \, \theta \in [0,1] \}$ be the class of all Bernoulli distributions.
We propose to focus on the family of models
$\mathcal{M}^{(n)} = \{P_\theta \,|\, \theta = \frac{i}{n},\; 0 \leq i \leq n\}$ that are models of description for finite size data samples.
$\mathcal{M}^{(n)}$ is related to the set of all the possible maximum likelihood estimates of $\theta$ (from $\mathcal{M}$) for binary strings of size $n$.
The interest of using $\mathcal{M}^{(n)}$ is that the number of model parameters is now finite instead of uncountable infinite.
Using a uniform prior on the model parameters in $\mathcal{M}^{(n)}$, we get $P(\theta = \frac{i}{n}) = 1/{|\mathcal{M}^{(n)}|}$, leading to $L(\theta) = \log (n+1)$.

Given $\theta = \frac{i}{n} \in \mathcal{M}^{(n)}$, we now have to encode the data $x^n$.

If $k(x^n)/n  \neq \theta$, we cannot encode the data and $P(x^n|\theta) = 0$.

If $k(x^n)/n  = \theta$, the observed data is consistent with the model parameter, and we assume that all the possible observable data are uniformly distributed.
The number of binary strings with $k$ ones is the binomial coefficient ${{n}\choose{k}}$. Thus the probability of observing one of them is $P(x^n|\widehat{\theta}(x^n)) = 1/{{n}\choose{k}}$.
We have a discrete likelihood that concentrates the probability mass on binary strings that can be observed given the model parameter. As a result, coding lengths are defined only for strings that are consistent with the model parameter.
This gives a total code length of

\begin{equation}
L(\widehat{\theta}(x^n), x^n) = \log(n+1) + \log \frac{n!} {k! (n-k)!},
\end{equation}
defined only when $\theta = \widehat{\theta}(x^n)$.

\paragraph{Generative model for the enumerative Bernoulli distribution.}
Given a sequence length $n$ and $\theta = \frac{i}{n} \in \mathcal{M}^{(n)}$, we can formulate these models as generative models of sequences with exactly $i$ ones and $n-i$ zeros. For example, from a sequence of $n$ zeros, we randomly choose $i$ times without replacement a zero in the sequence and replace it with a one.
For this generative model, we have the following likelihood, as seen previously:
\begin{equation}
P(x^n|\theta = \frac{i}{n}) = {\mathbb{1}_{\left\{ \frac{i}{n} = \frac{k(x^n)}{n}\right\}}} 1/{{n}\choose{k(x^n)}}.
\end{equation}
For the case of the Bayes mixture model with uniform prior $w(\theta) = \frac{1}{n+1}, \theta = \frac{i}{n},\; 0 \leq i \leq n$,
we have 
\begin{eqnarray}
P_{Bayes}(x^n) &=& \sum_{i=0}^n{w(\frac{i}{n}) P(x^n|\theta = \frac{i}{n})}, \\
 &=& \frac{1}{n+1} \frac{k(x^n)! (n-k(x^n))!} {n!} .
\end{eqnarray}
The negative log of this probability actually corresponds to the code length of the enumerative code.
Interestingly, the standard Bernoulli model and the enumerative one are related to slightly different generative models, but their Bayes mixture under the uniform prior leads to the same distribution.
In Section~\ref{NMLinterpretation}, we will see that on the opposite, their normalized maximum likelihood distribution is not the same.

\paragraph{Cardinality of models spaces.}
Let us consider the union of the $\mathcal{M}^{(n)}$ models for all the sample sizes:
\begin{equation}
\mathcal{M}^{(\mathbb{N})} = \cup_{n \in \mathbb{N}}{\mathcal{M}^{(n)}}.
\end{equation}
Interestingly, $\mathcal{M}^{(\mathbb{N})}$ is very close to $\mathcal{M}$, with $\theta \in \mathbb{Q}$ rather than $\theta \in \mathbb{R}$. Thus, the number of model parameters in $\mathcal{M}^{(\mathbb{N})}$ is countable infinite rather than uncountable infinite, which provides a significant simplification.

\subsection{NML interpretation}
\label{NMLinterpretation}

Let us compute the NML parametric complexity of this enumerative code, on the basis of the discrete likelihood presented in Section~\ref{secEnumBayesian}.
We have

\begin{eqnarray}
COMP^{(n)}(\theta) &=& \log \sum_{y^n \in X^n} {P_{\widehat{\theta}(y^n)}(y^n)}, \\
  &=& \log \sum_{k=0}^n {{{n}\choose{k}} \left({1} /{{n}\choose{k}}\right)},\\
	&=& \log (n+1).
\end{eqnarray}

Interestingly, we find exactly the same complexity term $\log (n+1)$ as the coding length of the best hypothesis in the enumerative two-part crude MDL code presented in Section~\ref{secEnumerativeB}.
This shows that the enumerative code is both a two-part and a one-part code. It is parametrization invariant and optimal w.r.t. the NML approach, with minimax regret guarantee.
Surprisingly, its parametric complexity is asymptotically twice that of the NML code described in Section~\ref{secNML}.
We further investigate on the comparison between the enumerative and NML codes in next section

\section{Code comparison for the Bernoulli distribution}
\label{secComparison}

In this section, we compare the NML code (Section~\ref{secNML}) and enumerative two-part crude MDL codes (Section~\ref{secRevisitedEnumerative}) for the Bernoulli distribution.

\subsection{Notation}

Let us use the names \emph{simplistic}, \emph{NML} and \emph{enumerative} for the specific MDL codes presented in Sections \ref{secSimplistic}, \ref{secNML} and \ref{secEnumerativeB}.
We also consider the \emph{random} code as a baseline: it corresponds to a direct encoding of each binary string $x^n$ with a coding length of $n \log 2$. The likelihood of each string $x^n$ is $1/2^n$, and as $\sum_{k=0}^n {{{n}\choose{k}}  1/2^n} = 1$, we have $COMP_{random}^{(n)}(\emptyset) = 0$ and 
$L_{random}\left(x^n|\emptyset \right) = n \log 2$.

Table~\ref{tableCodes} reminds the parametric and stochastic complexity of each considered code.

\begin{table}[htbp!]
\caption{Parametric and stochastic complexity per code.}
\label{tableCodes}
\centering
\renewcommand{\arraystretch}{1.5}
\begin{tabular}[10pt]{ccc}\hline
Code name         & $COMP_{name}^{(n)}$ & $L_{name}\left(x^n|\widehat{\theta}(x^n) \right)$ \\\hline 
\emph{enumerative}      & $\log (n+1)$    & $\log \frac {n!} {k! (n-k)!}$ \\
\emph{NML}  & $\frac{1}{2} \log \frac{n \pi}{2} + o(1)$ & $\log \frac {n^n} {k^k (n-k)^{n-k}}$ \\
\emph{simplistic} & $\log (n+1)$    & $\log \frac {n^n} {k^k (n-k)^{n-k}}$ \\
\emph{random}     & $0$    & $n \log 2$ \\\hline
\end{tabular}
\end{table}

As for the simplistic code, the coding length of the parameter is presented in place of the parametric complexity. The total coding length of the simplistic code has an overhead of about $\frac{1}{2} \log n$ compared to the NML code.
This confirms that the simplistic code is dominated by the NML code, as expected.

\subsection{Stochastic complexity term}

The stochastic complexity term of the enumerative code is always smaller than that of the NML code for non-degenerated binary strings:

\begin{equation}
\label{sc_Bernoulli}
	\forall n, \forall x^n \in X^n, \; 0 <k(x^n) < n, \;
	L_{enum}\left(x^n|\widehat{\theta}(x^n)\right) < L_{nml}\left(x^n|\widehat{\theta}(x^n)\right).
\end{equation}

An intuitive proof relies on the fact that the enumerative MDL likelihood assigns the same probability to all binary strings having the same number of ones, with a null probability for all the other strings. The NML likelihood also assigns the same probability to all binary strings having the same number of ones, but with a non-null probability for the other strings. Then they have to share a smaller probability mass, resulting in a smaller probability per string and a strictly greater coding length.

To gain further insights, let us approximate the difference of coding length:
$$\delta L\left(x^n|\widehat{\theta}(x^n)\right) = L_{nml}\left(x^n|\widehat{\theta}(x^n)\right) -L_{enum}\left(x^n|\widehat{\theta}(x^n)\right).$$
Using the approximation given in \citep{Grunwald07} (formula~4.36) with the Bernoulli parameter $\theta = \widehat{\theta}(x^n)$ , we have 

\begin{eqnarray}
L_{enum}\left(x^n|\theta\right) &=& \log {{n}\choose{\theta n}},\\
  &=& n H(\theta) - \log \sqrt{2 \pi n \mathrm{var} (\theta)} + O(1/n),\\
	&=& L_{nml}\left(x^n|\theta\right) -\frac{1}{2} \log (2 \pi n \mathrm{var}(\theta)) + O(1/n).
\end{eqnarray}

We get

\begin{equation}
\label{deltaL}
\delta L\left(x^n|\widehat{\theta}(x^n)\right) = \frac{1}{2} \log (2 \pi n \mathrm{var}(\theta)) + O(1/n).
\end{equation}

The difference of coding length is always positive but not uniform:
\begin{itemize}
	\item for $k(x^n) = 0$, $\delta L\left(x^n|\widehat{\theta}(x^n)\right) = 0$,
	\item for $k(x^n) \approx n/2$, $\delta L\left(x^n|\widehat{\theta}(x^n)\right) \approx \frac{1}{2} \log (\frac {n \pi}{2})$.
\end{itemize}

\medskip
These results demonstrate that the enumerative code provides a better encoding of the data with the help of the model for any binary strings, all the more for strings with equidistributed zeros and ones. The gain in coding length compared to the NML code grows as the logarithm of the sample size.

\subsection{Parametric complexity term}

Using inequality~\ref{sc_Bernoulli} and as the parametric complexity of code is the sum of the stochastic complexity over all possible strings, we get:

\begin{equation}
\label{pc_Bernoulli}
	\forall n > 1, COMP_{enum}^{(n)} > COMP_{NML}^{(n)}.
\end{equation}

Both terms are equal for $n=1$ and asymptotically, the parametric complexity of the enumerative code is twice that of the NML code (see Table~\ref{tableCodes}).

\begin{figure}[!htb]
\begin{center}
 \includegraphics[width=0.85\textwidth]{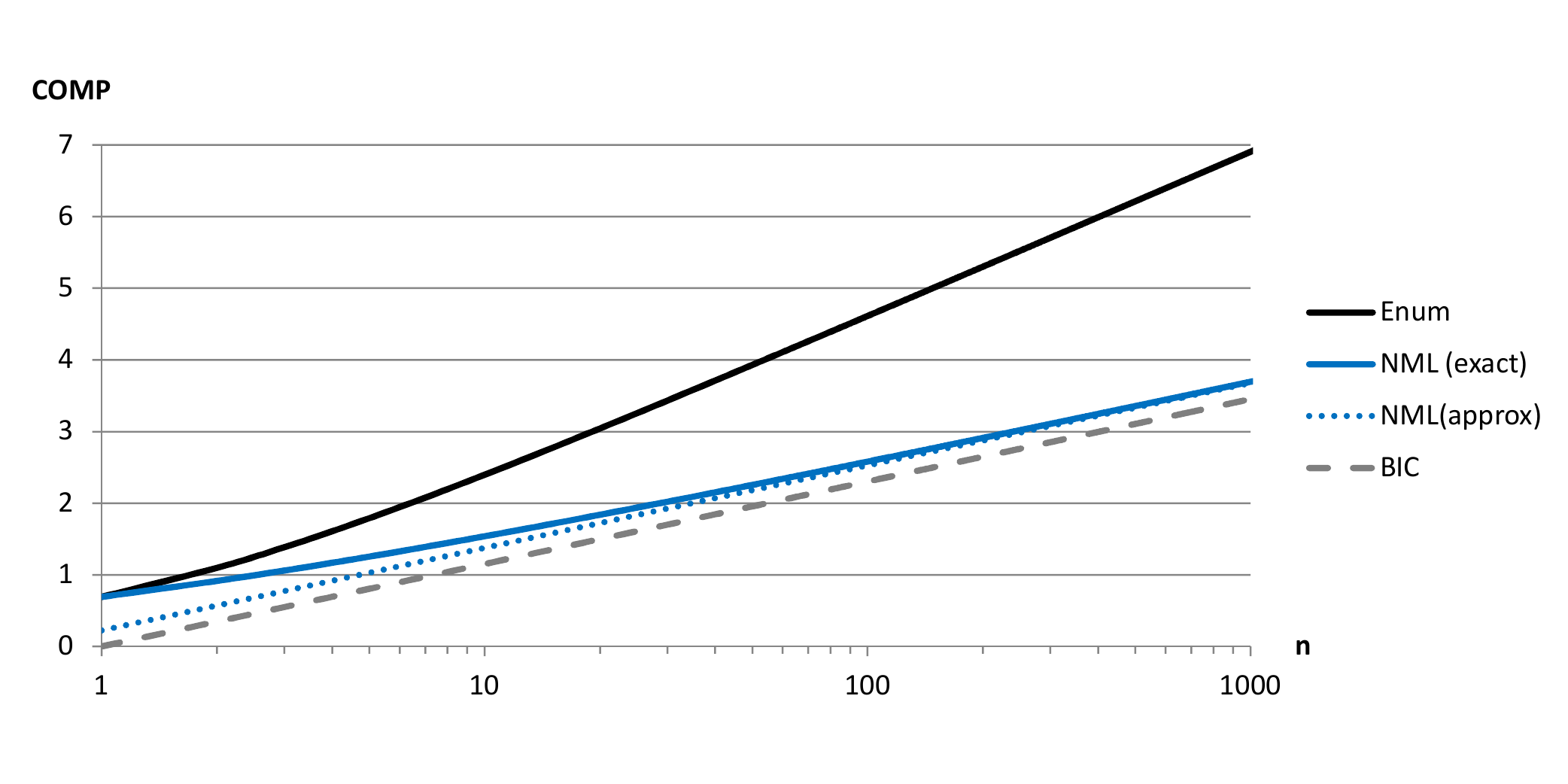}
 \caption{Parametric complexity for the Bernoulli model.}
 \label{fig_comp}
\end{center}
\end{figure}

We now focus on the non-asymptotic behavior of the parametric complexity terms and their approximations.
Figure~\ref{fig_comp} shows the value of the parametric complexity of the Bernoulli model, using the enumerative code, the NML code (exact numerical computation and approximation), as well as the related BIC penalization term.

The approximation of the NML parametric complexity is very good as soon as $n$ is beyond 100, but less accurate for small sample sizes.
Asymptotically, the parametric complexity of the enumerative code is twice that of the NML approach. It is always greater, but for very small sample sizes, the difference becomes smaller and smaller.

\subsection{Overall code length}

Both the enumerative and NML codes exploit universal distributions on all binary strings $x^n \in X^n$.
The compression of the data with the help of the model is better for the enumerative distribution, at the expense of a worse parametric complexity.
Adding the parametric complexity and stochastic complexity terms, previous sections show that the NML code is much shorter for degenerated binary strings:

\begin{eqnarray}
\mathrm{for} \; k(x^n) = 0 \;\, \mathrm{or} \;\, k(x^n) = n, \quad \quad&&\\ \nonumber
L_{enum}\left(\widehat{\theta}(x^n), x^n\right) &\approx& \log n\\ \nonumber
L_{nml}\left(\widehat{\theta}(x^n), x^n\right) &\approx& \frac{1}{2} \log n + \frac{1}{2} \log \frac{\pi}{2},
\end{eqnarray}

 whereas the enumerative code is slightly shorter for equidistributed binary strings (where $H(\widehat{\theta}(x^n)) \approx \log 2$), with a margin of $\log \frac{\pi}{2}$:

\begin{eqnarray}
\label{eqn_mixture}
\mathrm{for} \; k(x^n) \approx n/2, \quad \quad&&\\ \nonumber
L_{enum}\left(\widehat{\theta}(x^n), x^n\right) &\approx& \frac{1}{2} \log n - \frac{1}{2} \log \frac{\pi}{2} + n \log 2,\\ \nonumber
L_{nml}\left(\widehat{\theta}(x^n), x^n\right) &\approx& \frac{1}{2} \log n + \frac{1}{2} \log \frac{\pi}{2} + n \log 2.
\end{eqnarray}

\subsection{Expectation of the coding length of all binary strings}

\begin{figure}[!htb]
\begin{center}
 \includegraphics[width=0.85\textwidth]{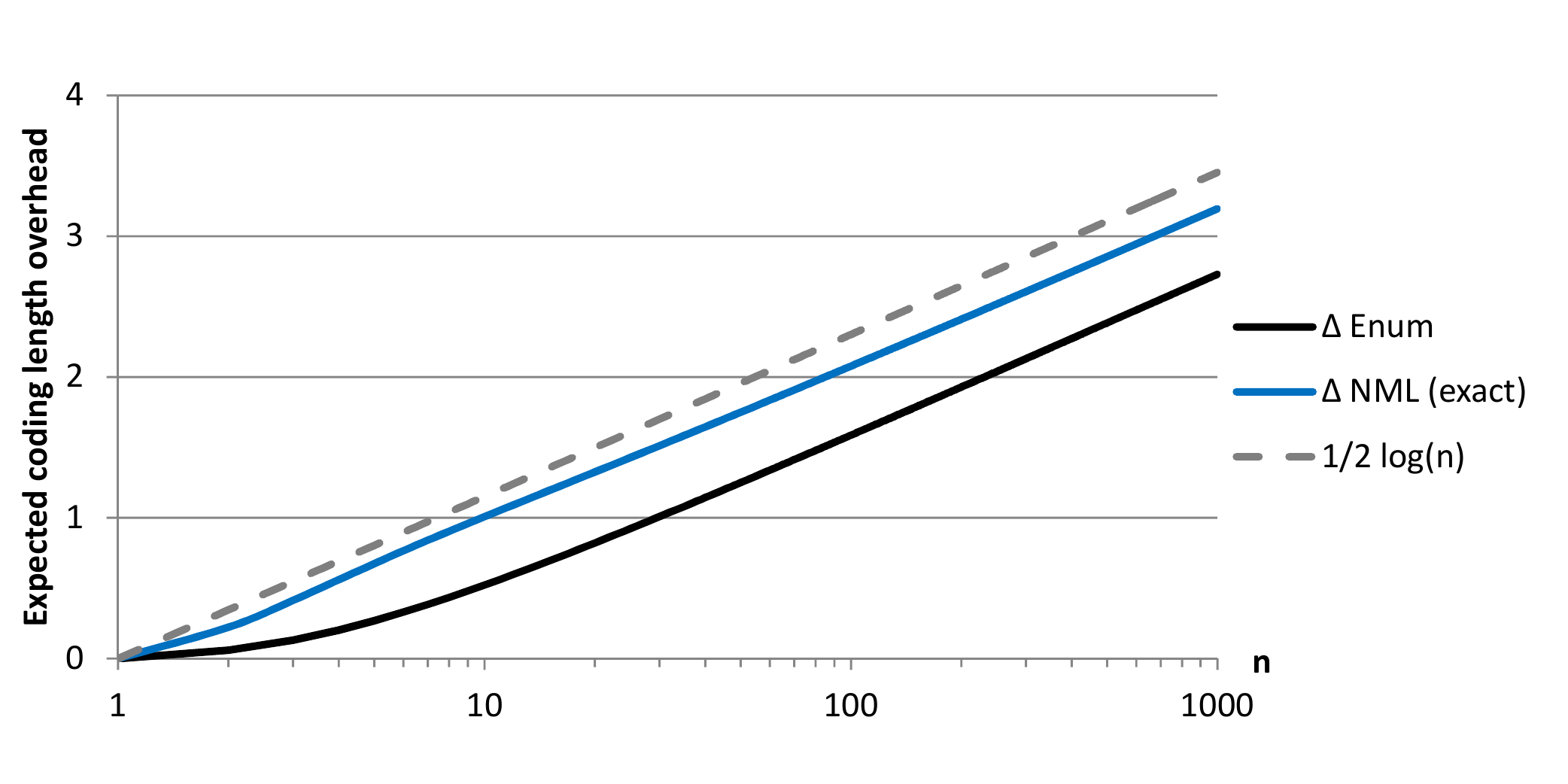}
 \caption{Expected overhead of coding length w.r.t. random model.}
 \label{fig_codinglength}
\end{center}
\end{figure}

Let us now estimate the expectation of the coding length for all binary strings under the uniform distribution, where $\forall x^n \in X^n, p(x^n) = 1/{2^n}$.
\begin{eqnarray}
\label{eqn_codinglength}
\mathrm{E}\left(L\left(\widehat{\theta}(x^n), x^n\right)\right) &=& \frac{1} {2^n} \sum_{x^n \in X^n} {L\left(\widehat{\theta}(x^n), x^n\right)},\\ \nonumber
  &=& \frac{1} {2^n} \sum_{k=0}^n {{{n}\choose{k}} L\left(\widehat{\theta}(x^n), x^n\right)}.
\end{eqnarray}

We perform an exact numerical calculation using the exact value of the NML model complexity term, for all $n, 1 \leq n \leq 1000$.
Figure~\ref{fig_codinglength} reports the expected coding length for the enumerative and NML codes minus that of the random code ($n \log 2$).
The results show that both codes have an average overhead of about $1/2 \log n$ compared to the direct encoding of the binary strings, and that, 
under the uniform distribution, 
the enumerative code always compresses the data better on average that the NML code, especially in the non-asymptotic case.

Actually, averaging on all binary strings is the same as considering exhaustively all the binary string outcomes of a Bernoulli distribution with parameter $\theta = 1/2$.
Using the central limit theorem, the proportion of binary strings $x^n$ where $k(x^n)/n \approx 1/2$ goes to 1 as $n$ goes to infinity, which explains why the shorter coding lengths obtained with the enumerative code for binary strings with $k(x^n)/n \approx 1/2$ provide the main contribution in the expectation.
Using Formula~\ref{eqn_mixture}, the expected coding length of the enumerative code is asymptotically better than that of the NML code by a margin of $\log \frac{\pi}{2}$.

\subsection{Percentage of compressible binary strings}
\label{percentCompressibleBernoulli}

We now focus on the percentage $p_{compressible}$ of compressible binary strings using both the enumerative and NML codes, that is the percentage of binary strings with coding length shorter than $n \log 2$:

\begin{eqnarray}
\label{eqn_percentcompressible}
p_{compressible} &=& \frac{1} {2^n} \sum_{x^n \in X^n}
 {\mathbb{1}_{\left\{L\left(\widehat{\theta}(x^n), x^n\right) \leq n \log 2\right\}}},\\ \nonumber
  &=& \frac{1} {2^n} \sum_{k=0}^n {{{n}\choose{k}} 
	\mathbb{1}_{\left\{L\left(\widehat{\theta}(x^n), x^n\right) \leq n \log 2\right\}}}.
\end{eqnarray}

\begin{figure}[!htb]
\begin{center}
 \includegraphics[width=0.85\textwidth]{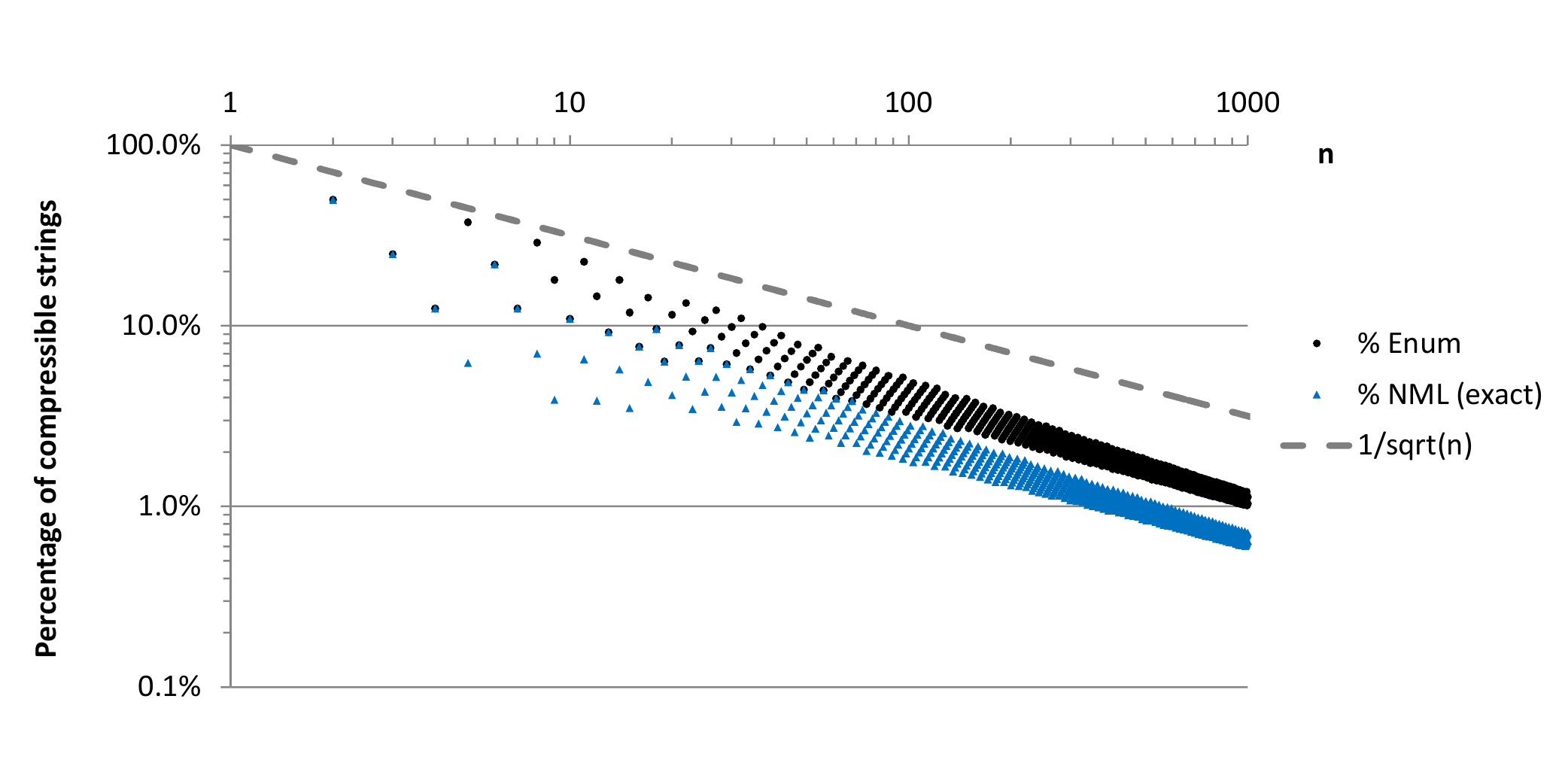}
 \caption{Percentage of compressible binary strings.}
 \label{fig_percentcompressible}
\end{center}
\end{figure}

As previously, we perform an exact numerical calculation for all $n, 1 \leq n \leq 1000$.
Figure~\ref{fig_percentcompressible} shows that the percentage of compressible strings decreases at a rate of $O(1/\sqrt{n})$ for both codes, as expected.
However, the enumerative code always compresses more binary strings than the NML code. Due to the discrete decision threshold in formula~\ref{eqn_percentcompressible}, the exact computed percentage values are not smooth like in Figure~\ref{fig_codinglength}, especially in the non-asymptotic case for small string sizes, but the overall tendency appears clearly for large sample sizes.
In the asymptotic case, around $60\%$ more strings can be compressed using the enumerative code (empirical evaluation).

\subsection{Distribution of compression rates}
\label{compressionrateBernoulli}

We now focus on the distribution of the compression rates, that is the ratio of the coding length of a binary string using the Bernoulli versus the random model:

\begin{equation}
\label{eqn_compressionrate}
\%compression = \frac{L\left(\widehat{\theta}(x^n), x^n\right)} {n \log 2}.
\end{equation}

\begin{figure}[!htb]
\begin{center}
 \includegraphics[width=0.7\textwidth]{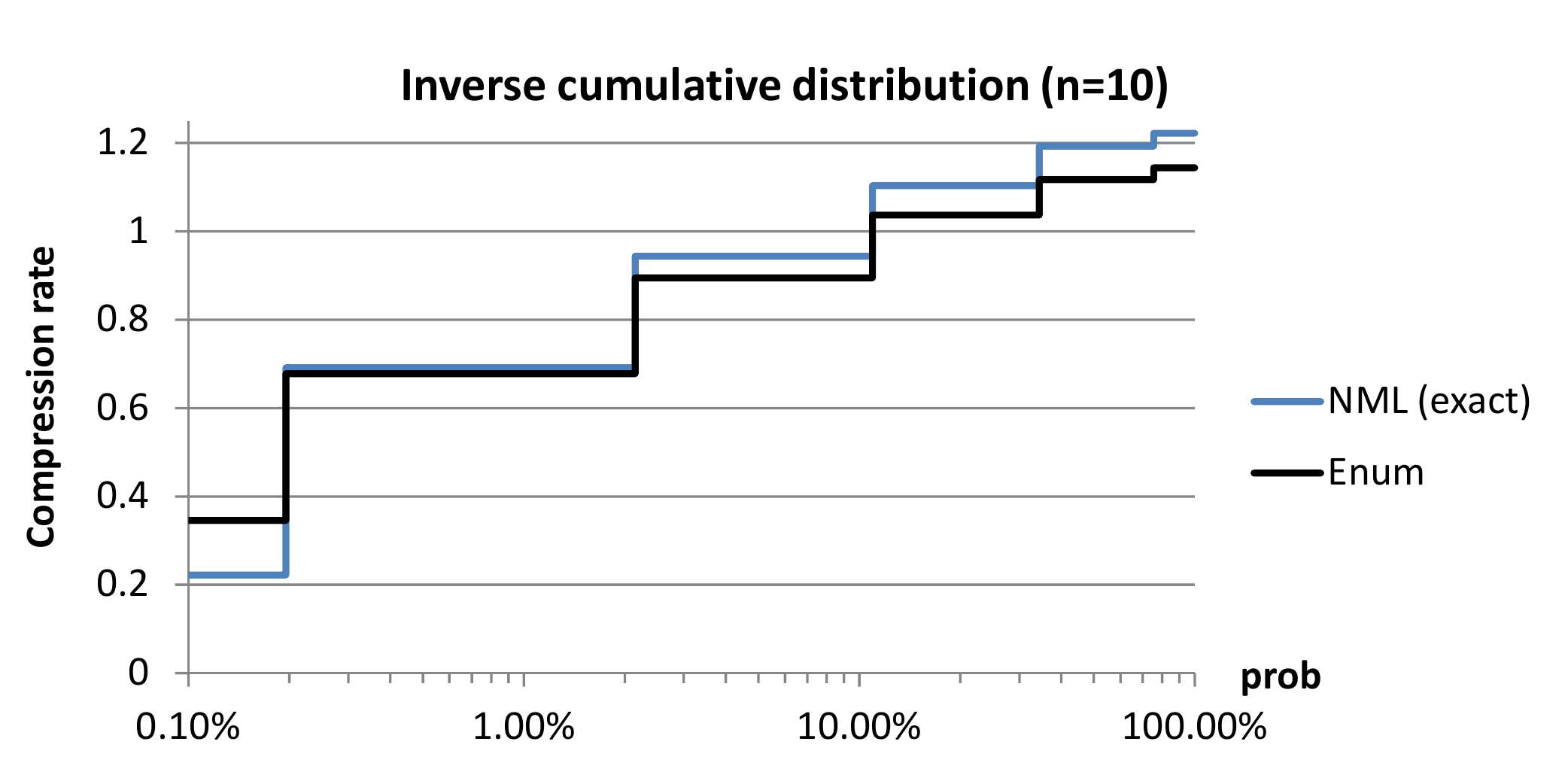}\\
 \includegraphics[width=0.7\textwidth]{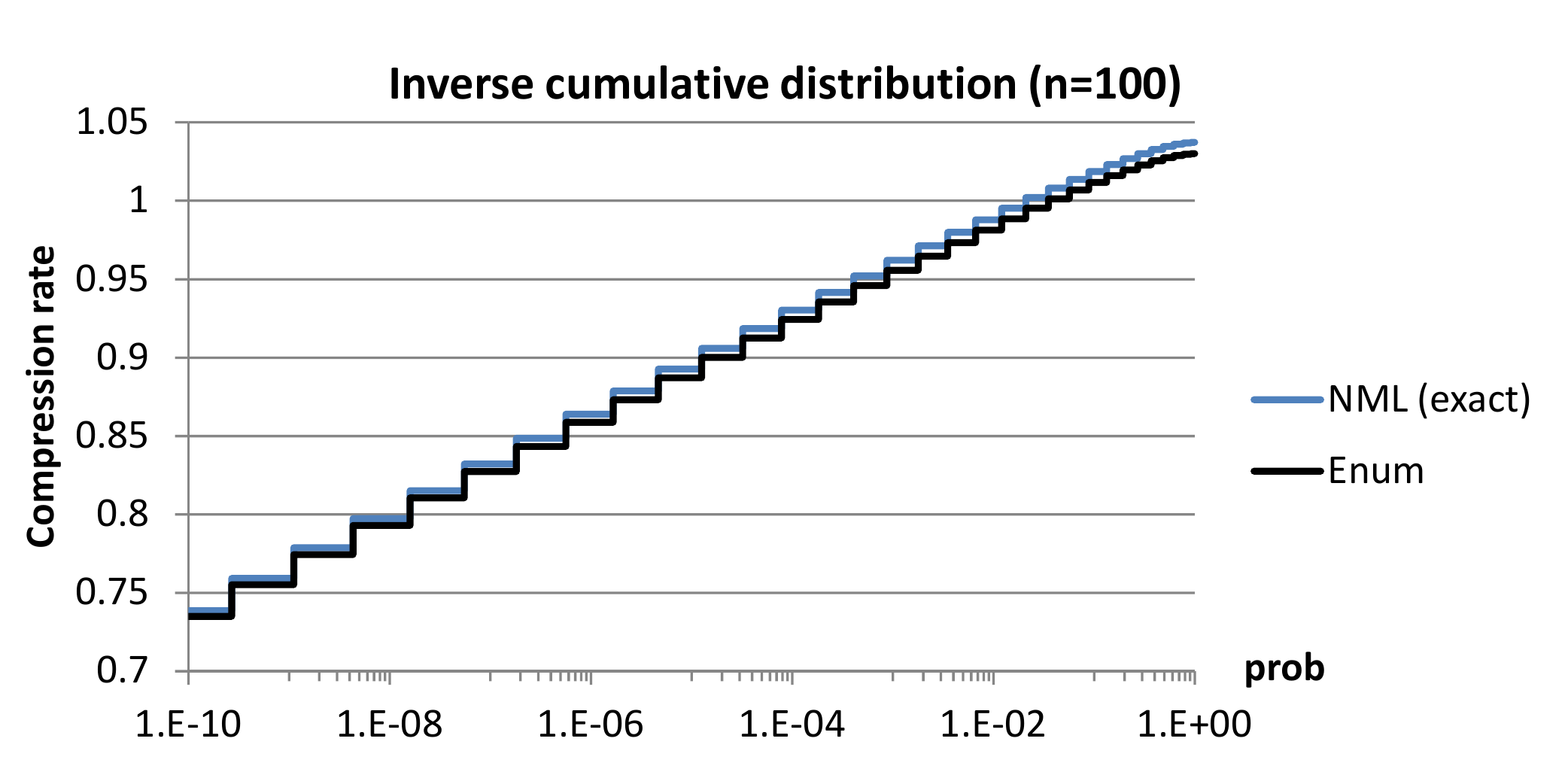}\\
 \includegraphics[width=0.7\textwidth]{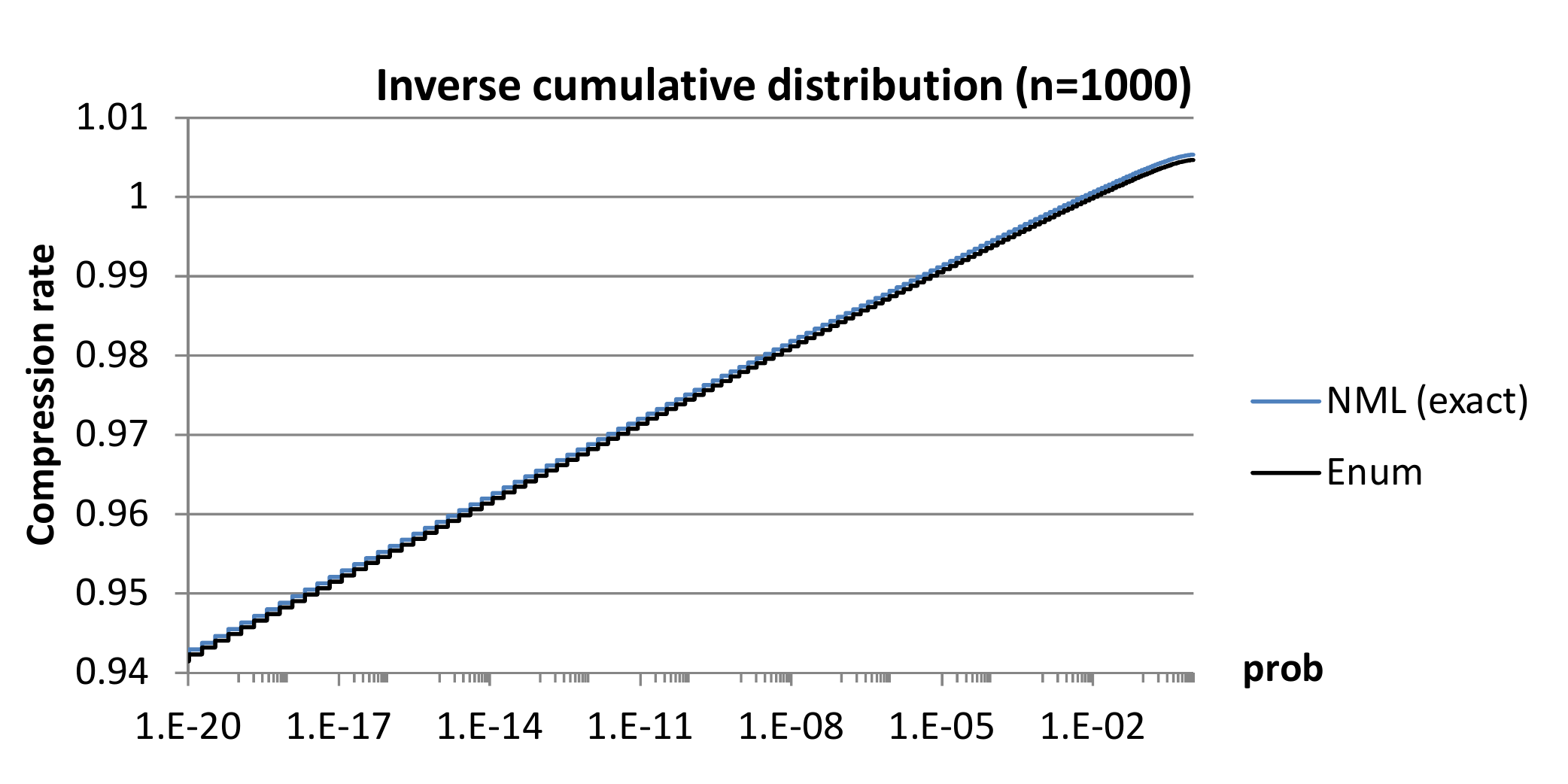}
 \caption{Inverse cumulative distribution of compression rates for strings of size 10, 100, 1000.}
 \label{fig_compressionrate}
\end{center}
\end{figure}

Figure~\ref{fig_compressionrate} shows the inverse cumulative distribution of compression rates for binary strings of size 10, 100, 1000, using the NML or enumerative code .
For example, from the 1024 ($2^{10}$) strings of size 10, only $0.2\%$ (the two ``pure'' strings) can be compressed better with the NML than with the enumerative code. 
All the other strings are better compressed with the enumerative code. For both codes, $11\%$ of the strings are compressible ($\%compression < 1$).
For string of size 100, only $3.0\; 10^{-15}\%$ of the strings are better compressed with the NML code.
However, $2.1\%$ of the strings are compressible using the NML code, which is less that the $3.5\%$ obtained using the enumerative code.
The tendency is the same for string of size 1000. A tiny portion of the strings, those with almost only zeros or only ones, are better compressed using the NML code.
All the other string are better compressed using the enumerative code, with difference growing for balanced strings. This results in a greater number of compressible strings using the enumerative code.

Let us evaluate the asymptotic value of the Bernoulli parameter $\theta$ for which both codes achieve the same compression rate.
Using Table~\ref{tableCodes} and Formula~\ref{deltaL} in the asymptotic case, we get :

\begin{eqnarray}
\label{eqn_compareCodes}
\delta\left(COMP^{(n)} + L\left(x^n|\widehat{\theta}(x^n)\right) \right) = 0 
 &\Leftrightarrow& \frac{1}{2} \log \frac{n \pi}{2} - \log(n+1) +\frac{1}{2} \log (2 \pi n \mathrm{var}(\theta)) = 0 \nonumber \\ 
&\Leftrightarrow& \log (2 \pi n \mathrm{var}(\theta)) = \log \frac {(n+1)^2}{n \pi /2}  \nonumber \\
&\Leftrightarrow& \mathrm{var}(\theta) = \frac {(n+1)^2}{n^2 \pi^2} \approx  \frac {1}{\pi^2} \nonumber \\
&\Leftrightarrow& \theta(1-\theta) = \frac {1}{\pi^2}
\end{eqnarray}

Equation~\ref{eqn_compareCodes} has two solutions: $\theta  = 1/2 (1 \pm \sqrt{1-4/{\pi^2}})$, that is $\theta \approx 0.114$ and $\theta \approx 0.886$. Thus asymptotically, the enumerative code better compresses the strings for $\theta \in [0.114, 0.886]$, that is around $77\%$ of the values of $\theta$.

Overall, both the NML and enumerative codes have the same asymptotic behavior, with tiny differences in compression rates. However, the enumerative code allows to better compress far more strings, both in the non-asymptotic and asymptotic cases.

\subsection{Detection of a biased coin}
\label{biasedCoin}

We apply the previous Bernoulli codes to the problem of detection of a biased coin.
A fair coin is a randomizing device with two states named \emph{heads} and \emph{tails} that are equally likely to occur. It can be modeled using a Bernoulli process with $\theta_{fair} = \frac{1}{2}$.
For a biased coin, the heads and tails are not equally likely to occurs, and the related Bernoulli parameter is $\theta_{bias} \neq \frac{1}{2}$.

The problem is to determine whether a coin is biased  given a limited sample of Bernoulli trials.
Given a sample $x^n$ of trials, we compute the coding length of this sample using either the NML or the enumerative code and decide that the coin is biased if its coding length is shorter than that of the random code ($n\log 2$). For a given size $n$ and a code (e.g. enumerative or NML), we compute the probability of detecting the biased coin by averaging the detection over all the possible samples of size $n$.
Formally, for each code, we thus compute:

\begin{eqnarray}
\label{eqn_biasedCoin}
prob^D (\theta_{bias}, n)
&=& \mathrm{E}_{B(\theta_{bias})} 
\left( \mathbb{1}_{ \left\{ L\left(\widehat{\theta}(x^n), x^n\right) < n \log 2 \right\} } \right) \\
&=& \sum_{k=0}^n {{{n}\choose{k}} \theta_{bias}^{k} (1-\theta_{bias})^{n-k} 
\mathbb{1}_{ \left\{ L\left(\widehat{\theta}(x^n), x^n\right) < n \log 2 \right\} } }.
\end{eqnarray}

\begin{figure}[!htb]
\begin{center}
 \includegraphics[width=0.7\textwidth]{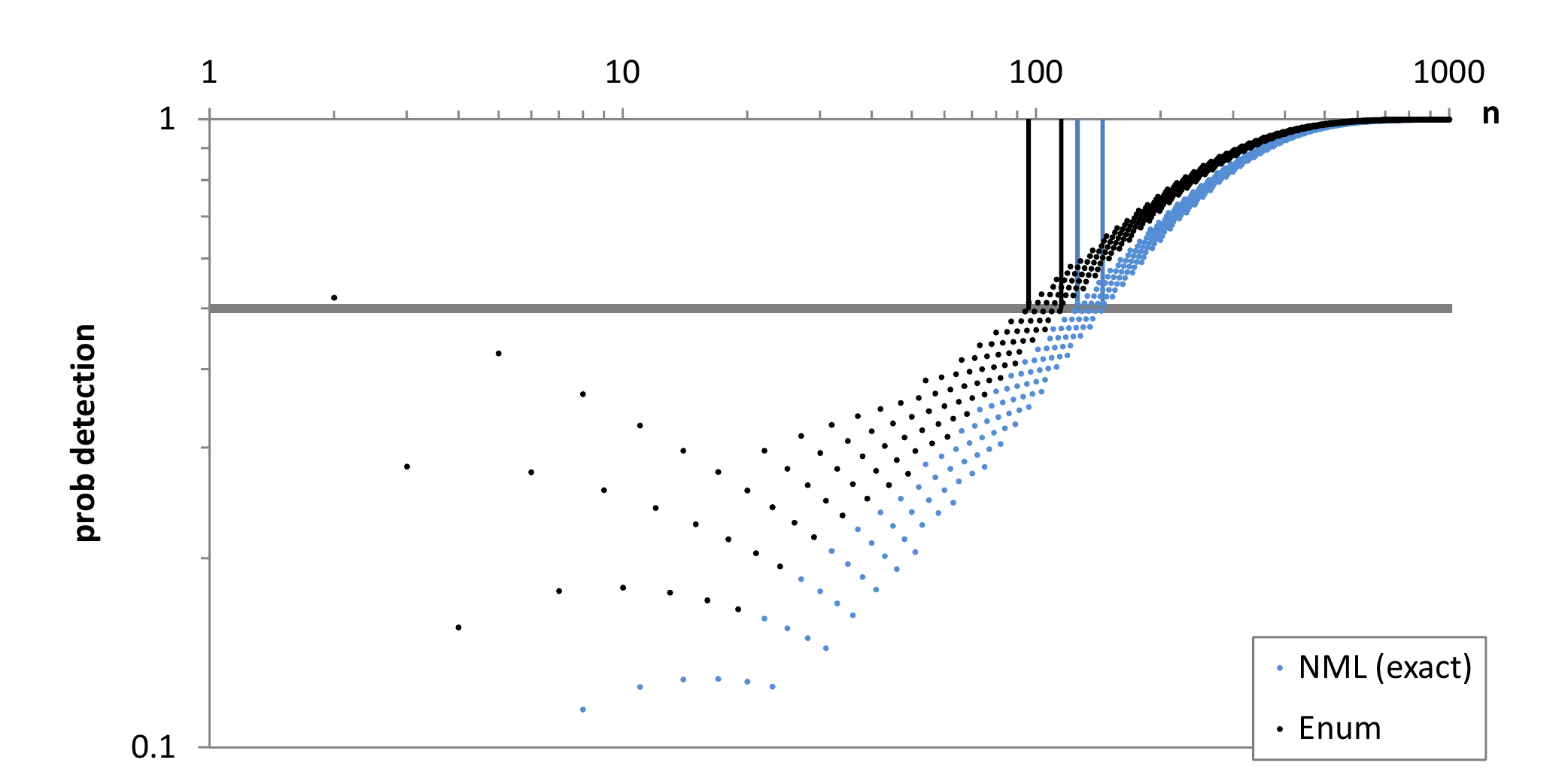}\\
 \caption{Probability of detection of of a biased coin where $\theta_{bias} = 0.40$.}
 \label{fig_BiasedCoinProbDetection}
\end{center}
\end{figure}

The issue is to be able to detect the biased coin with the minimum number of trials.
Using Formula~\ref{eqn_biasedCoin}, we then determine the first value of $n$ where the probability of detecting the biased coin is beyond $50\%$.
For example, Figure~\ref{fig_BiasedCoinProbDetection} shows the probability of detection of a biased coin ($\theta_{bias} = 0.40$) for sample size rangin from 1 to 1000, using the enumerative and NML codes.
The horizontal gray line represents a probability of $50\%$ of detecting the bias.
As Formula~\ref{eqn_biasedCoin} is not strictly increasing with $n$ and unstable for tiny $n$ (for reasons similar as in Section~\ref{percentCompressibleBernoulli}), we collect the two following lower and upper thresholds of sample sizes:

\begin{eqnarray}
\label{eqn_thresholdBiasedCoin}
\underline{n}^{\; D} (\theta_{bias})
&=& \min_{n \geq 10} \{prob^D (\theta_{bias}, n) \geq 50\% \},\\
\overline{n}^{\; D} (\theta_{bias})
&=& \max_{n \geq 10} \{prob^D (\theta_{bias}, n) \leq 50\% \}.
\end{eqnarray}

In Figure~\ref{fig_BiasedCoinProbDetection} for example, we have 
$\underline{n}^{\; D} (\theta_{bias}) = 96$ and $\overline{n}^{\; D} (\theta_{bias}) = 115$ for the enumerative code and
$\underline{n}^{\; D} (\theta_{bias}) = 126$ and $\overline{n}^{\; D} (\theta_{bias}) = 145$ for the NML code, that thus needs around $10\%$ more trials to detect the biased coin.

\begin{figure}[!htb]
\begin{center}
 \includegraphics[width=0.7\textwidth]{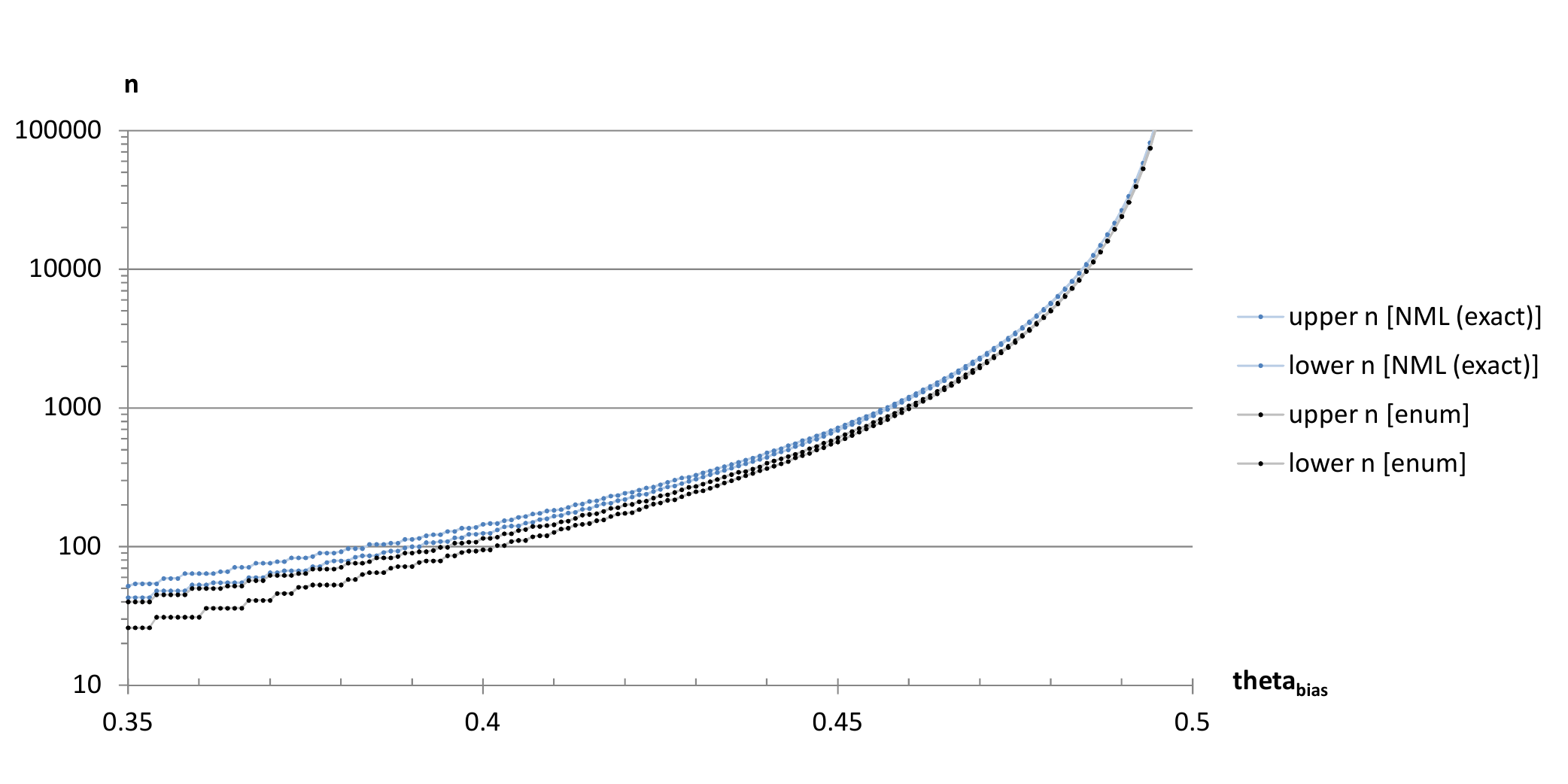}\\
 \caption{Minimum sample size to detect a biased coin with probability greater than $50\%$.}
 \label{fig_thresholdBiasedCoin}
\end{center}
\end{figure}

Figure~\ref{fig_thresholdBiasedCoin} shows the detection thresholds computed using the NML or enumerative codes for $\theta_{bias} \in [0.35; 0.5]$. 
As expected, the min sample size necessary to detect a biased coin increases quickly when $\theta_{bias}$ becomes close to $\frac{1}{2}$. For $\theta_{bias} \approx 0.46$, around 1,000 trials are necessary to detect the bias, and for $\theta_{bias} \approx 0.495$, around 100,000 trials are necessary.
Although all the thresholds are quite close, the enumerative code always needs smaller sample sizes to detect the biased coin.

\begin{figure}[!htb]
\begin{center}
 \includegraphics[width=0.7\textwidth]{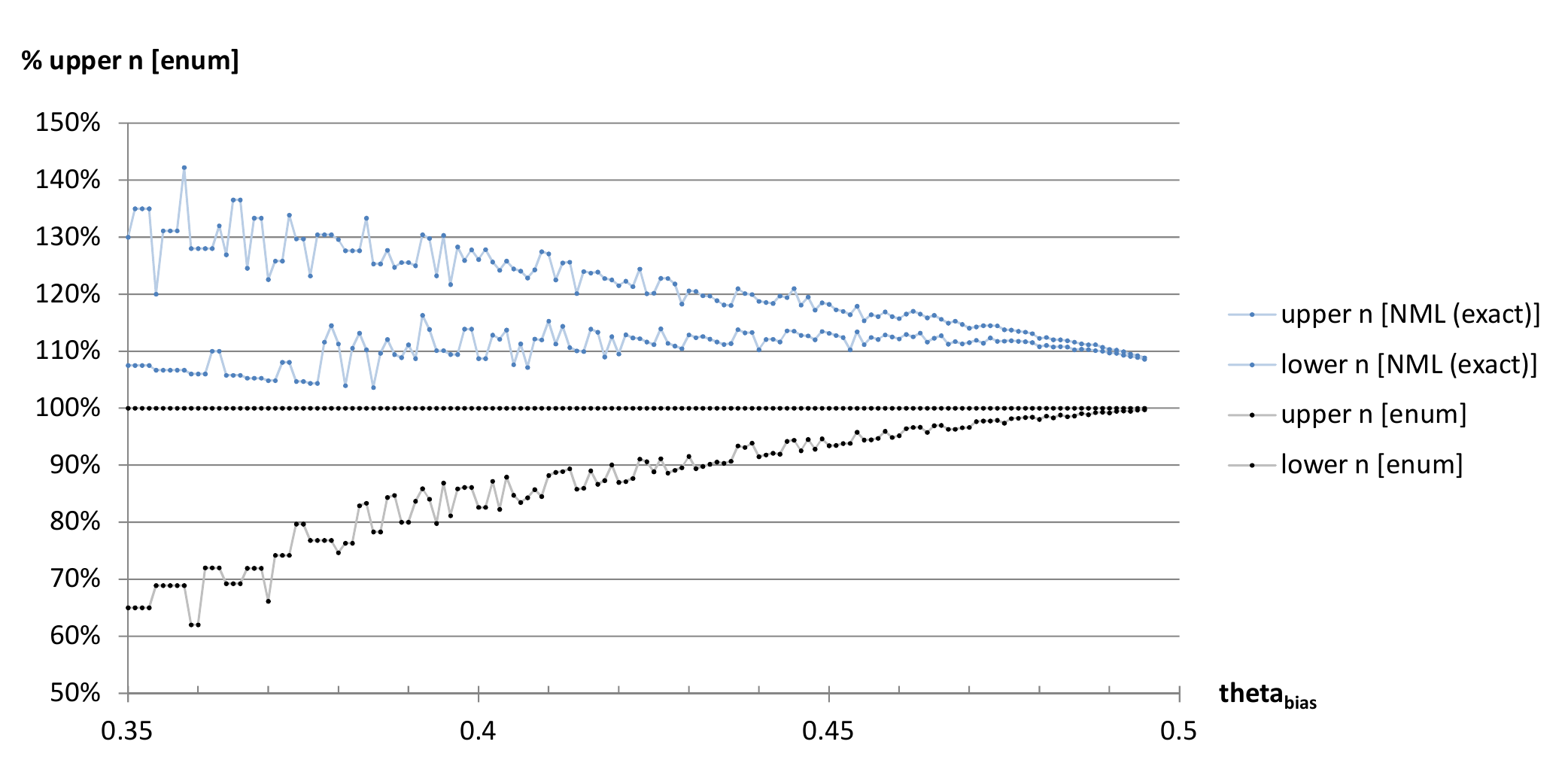}\\
 \caption{Minimum sample size to detect a biased coin with probability greater than $50\%$.}
 \label{fig_thresholdBiasedCoinRatio}
\end{center}
\end{figure}

To better compare the threshold without being hampered by the logarithmic scale of the sample size, Figure~\ref{fig_thresholdBiasedCoinRatio} shows all the detection thresholds normalized by the enumerative upper threshold. The lower and upper thresholds converge both for the enumerative and NML codes. However, the difference between code does not vanish with the sample size. At least withing the range explored in this experiment, up to 100,000 trials, the enumerative code always needs around $10\%$ less samples on average than the NML code to detect a biased coin.

\paragraph{False versus true positive rate.}

The probability of detecting a bias when a coin is actually biased can be interpreted as a true positive rate, and when the coin is fair as a false positive rate.
Given this, the enumerative code needs less samples than the NML code to detect a bias with a true positive rate greater than $50\%$.
In the case of a fair coin, the false positive rate of both codes decreases at a rate of O$(1/\sqrt n)$, as shown in the experiment related to the percentage of compressible strings (cf. Section~\ref{percentCompressibleBernoulli}: formula~\ref{eqn_percentcompressible} is the same as formula~\ref{eqn_thresholdBiasedCoin} for $\theta_{bias} = \frac{1}{2}$). Still, the false positive rate is about $60\%$ higher for the enumerative code than for the NML code. 

Overall, the enumerative code compresses most binary strings slightly better than the NML code, resulting in a better sensitivity to biased coins at the expense of more false detections in case of fair coins.

\subsection{Biased versus fair coin classification}
\label{secCoinClassification}

To further investigate on the comparison between the NML and enumerative codes, we suggest a classification experiment where the objective is to predict whether a coin if fair or biased. 
Let $\theta_{bias} \in [0;1]$ and $n \in \mathbb{N}^*$ be fixed parameters.
The instances to classify are sequences $x^n$ of $n$ trials generated with equal probability ($p_F=p_B = \frac{1}{2}$) either from a fair coin ($\theta = \frac{1}{2}$) or from a biased coin ($\theta = \theta_{bias}$).
The objective is to predict whether the coin that produced each sequence was fair or biased. 
As in Section~\ref{biasedCoin}, we evaluate both the NML and enumerative codes as classifiers by predicting a bias if they can encode a sequence with a coding length shorter than that of the random code ($n\log 2$), and predicting fair otherwise.

\begin{table}[htbp!]
\caption{Coin classification results.}
\label{coinContigencyTable}
\centering
\begin{tabular}[10pt]{|c|c|c|}\hline
{\scriptsize Real $\downarrow$ Predicted $\rightarrow$} & \qquad Bias \quad \quad & \qquad Fair \quad \quad \\\hline 
Bias & \qquad TP \quad \quad & \qquad FN \quad \quad \\\hline 
Fair   & \qquad FP \quad \quad & \qquad TN \quad \quad \\\hline
\end{tabular}
\end{table}

The result can be analyzed in terms of a contingency table, as illustrated in Table~\ref{coinContigencyTable}:
\begin{itemize}
	\item true positive (TP): detecting bias correctly,
	\item false positive (FP): detecting bias when there is none,
	\item true negative (TN): detecting fair correctly,
	\item false negative (FN): detecting fair when the coin is biased.
\end{itemize}

In this experiment, we focus on the correct detections:
\begin{itemize}
	\item sensitivity or true positive rate $TPR= TP/(TP+FN)$ for the correct detection of bias,
	\item specificity or true negative rate $TNR= TN/(TN+FP)$ for the correct detection of fair,
	\item accuracy $ACC = (TP+TN)/(TP+FP+FN+TN)$ for the global rate of correct detection.
\end{itemize}

For given $\theta_{bias}$ and $n$ parameters and for each code, we compute the expectation of the indicators by integrating other the distribution of all the sequences issued from the generation process.

\begin{eqnarray}
\label{eqn_TPR}
E(TPR)
&=& \mathrm{E}_{B(\theta_{bias})} 
\left( \mathbb{1}_{ \left\{ L\left(\widehat{\theta}(x^n), x^n\right) < n \log 2 \right\} } \right),\\
&=& \sum_{k=0}^n {{{n}\choose{k}} \theta_{bias}^{k} (1-\theta_{bias})^{n-k}
\mathbb{1}_{ \left\{ L\left(\widehat{\theta}(x^n), x^n\right) < n \log 2 \right\} } },\\
E(TNR)
&=& \mathrm{E}_{B(1/2)} 
\left( \mathbb{1}_{ \left\{ L\left(\widehat{\theta}(x^n), x^n\right) \geq n \log 2 \right\} } \right), \\
&=& \frac{1}{2^n} \sum_{k=0}^n {{{n}\choose{k}} 
\mathbb{1}_{ \left\{ L\left(\widehat{\theta}(x^n), x^n\right) \geq n \log 2 \right\} } },\\
E(ACC)
&=& p_B E(ETR) + p_F E(TNR),\\
&=& \frac{E(ETR) + E(TNR)}{2}.
\end{eqnarray}

\begin{figure}[!htb]
\begin{center}
\includegraphics[width=0.45\textwidth]{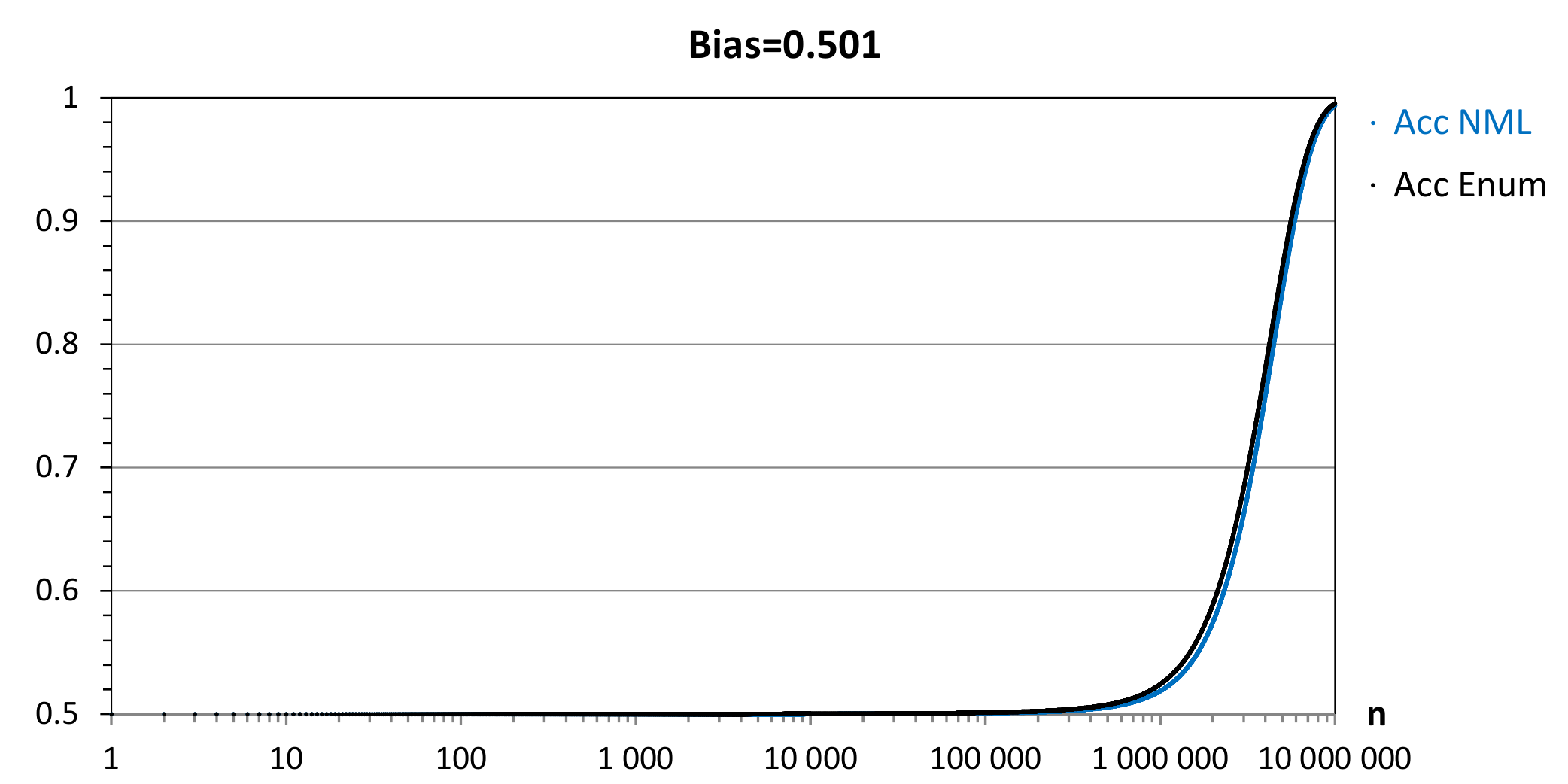}
\includegraphics[width=0.45\textwidth]{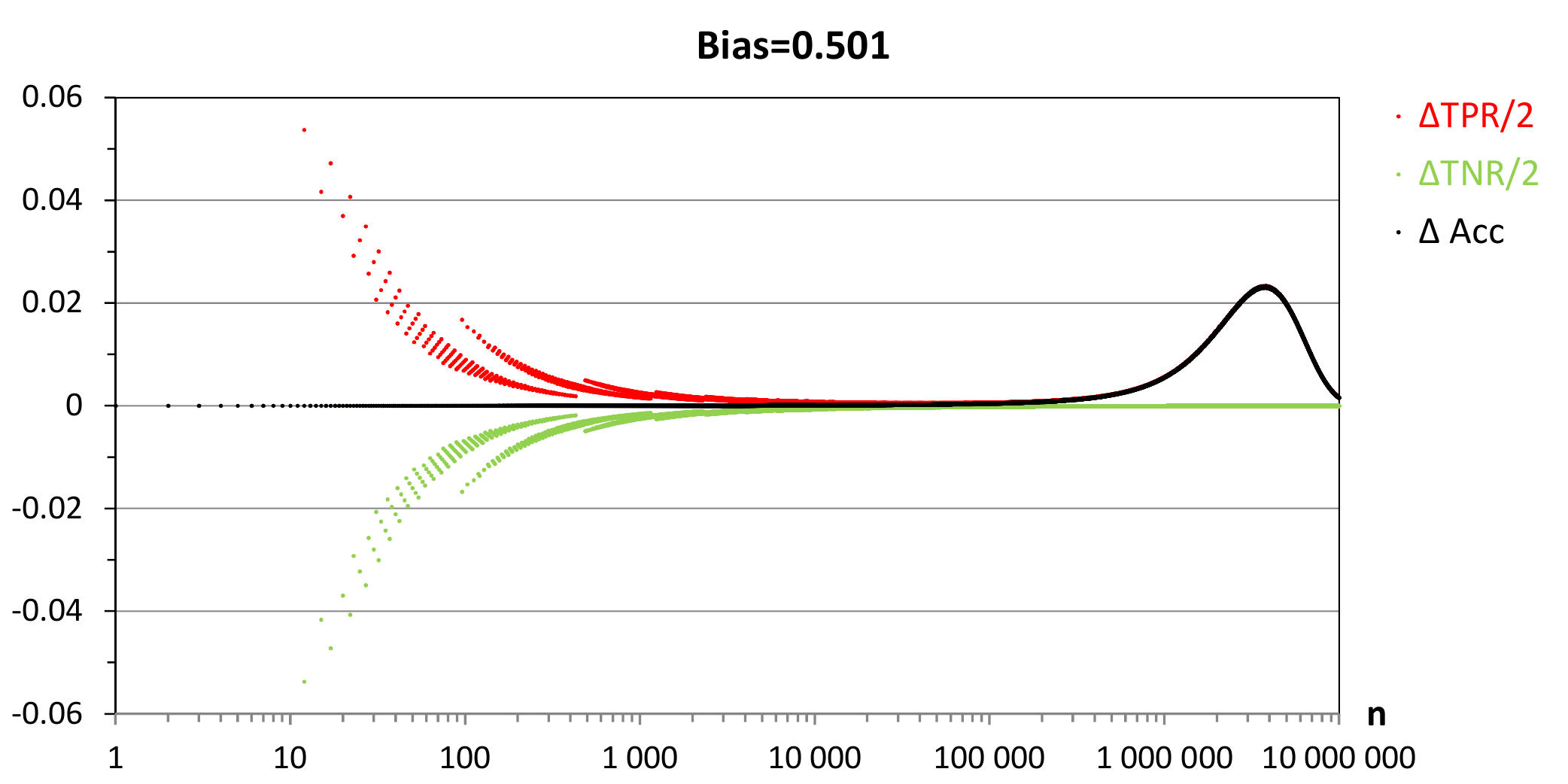}\\
\includegraphics[width=0.45\textwidth]{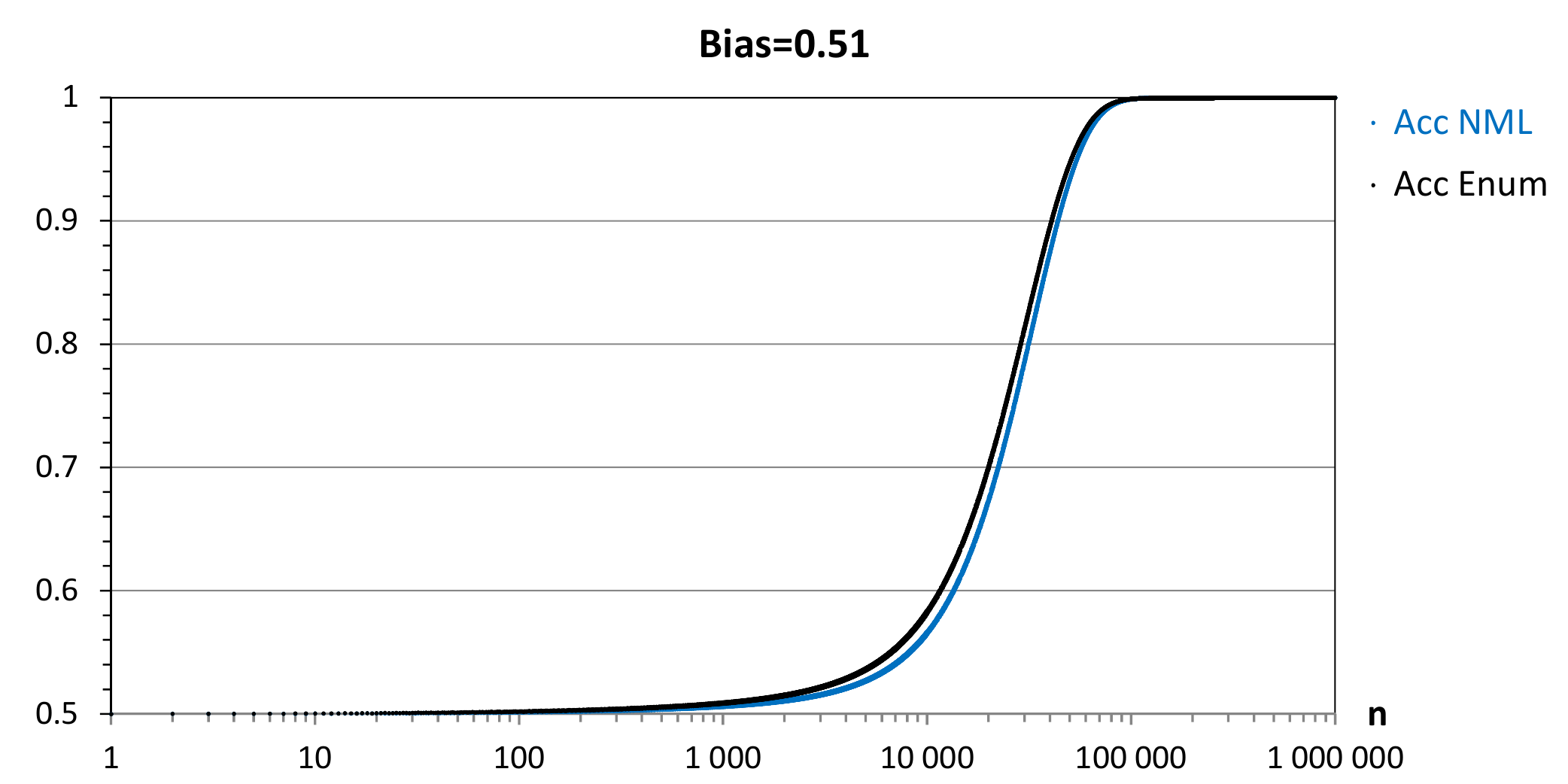}
\includegraphics[width=0.45\textwidth]{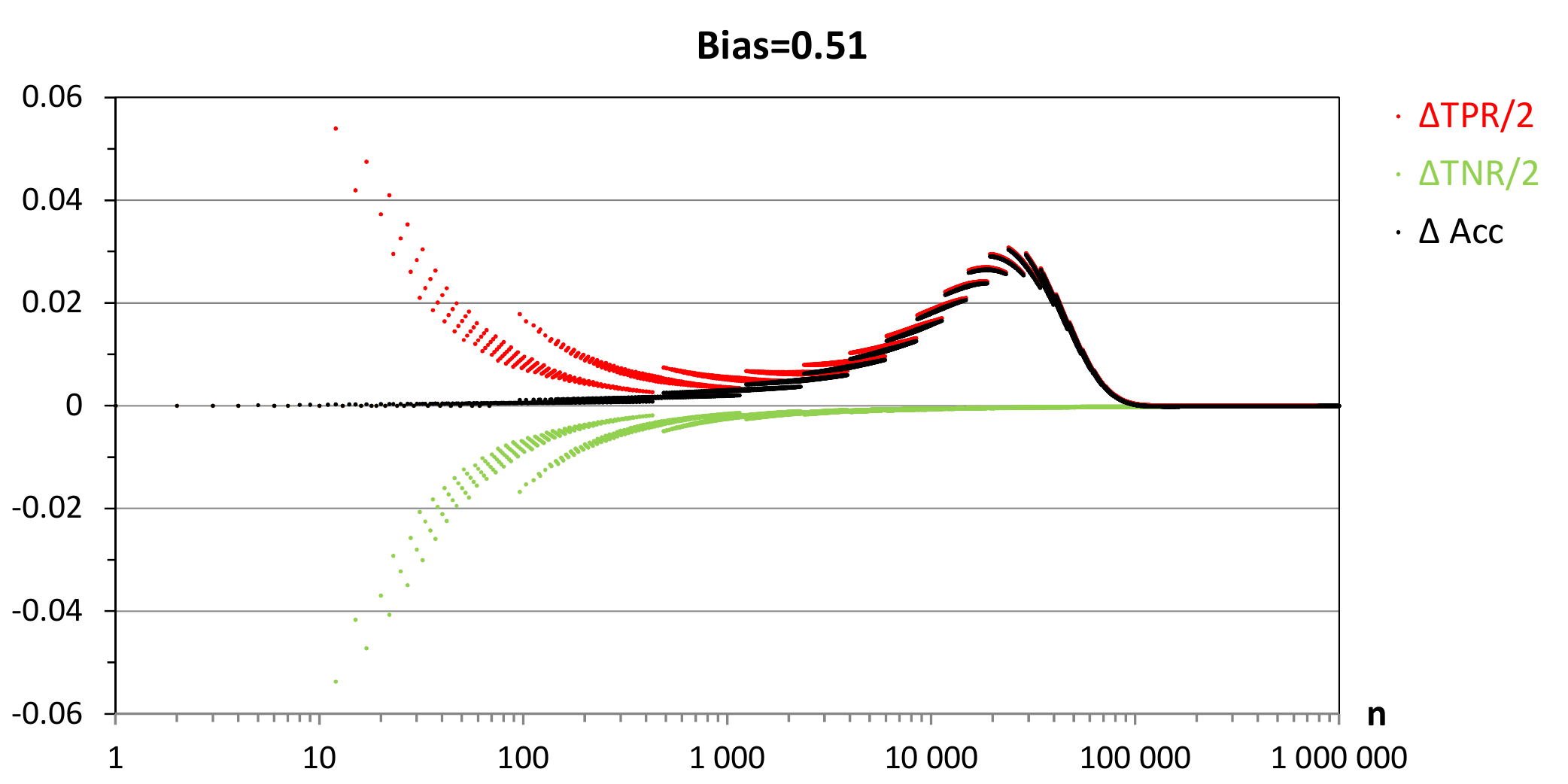}\\
\includegraphics[width=0.45\textwidth]{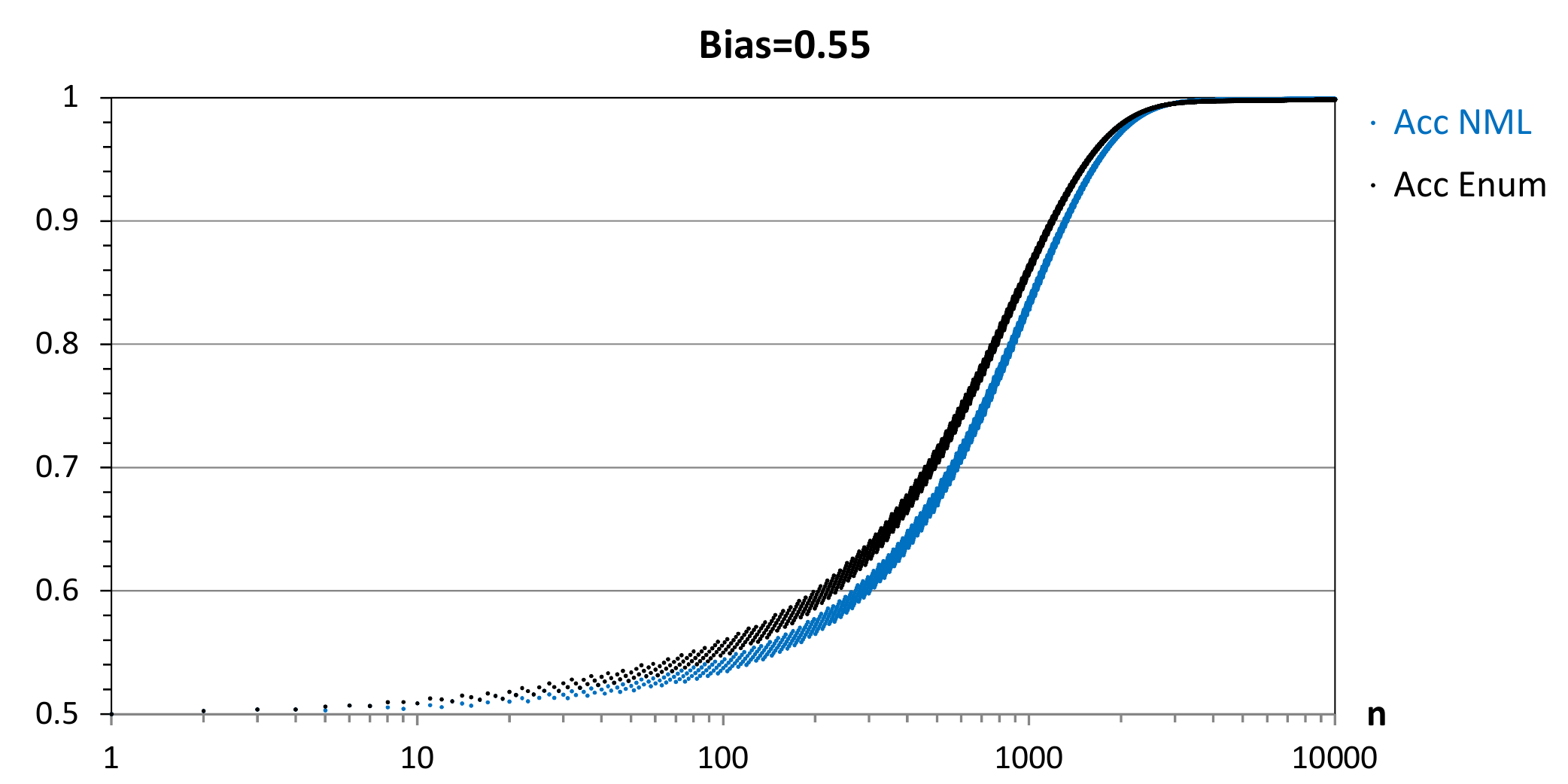}
\includegraphics[width=0.45\textwidth]{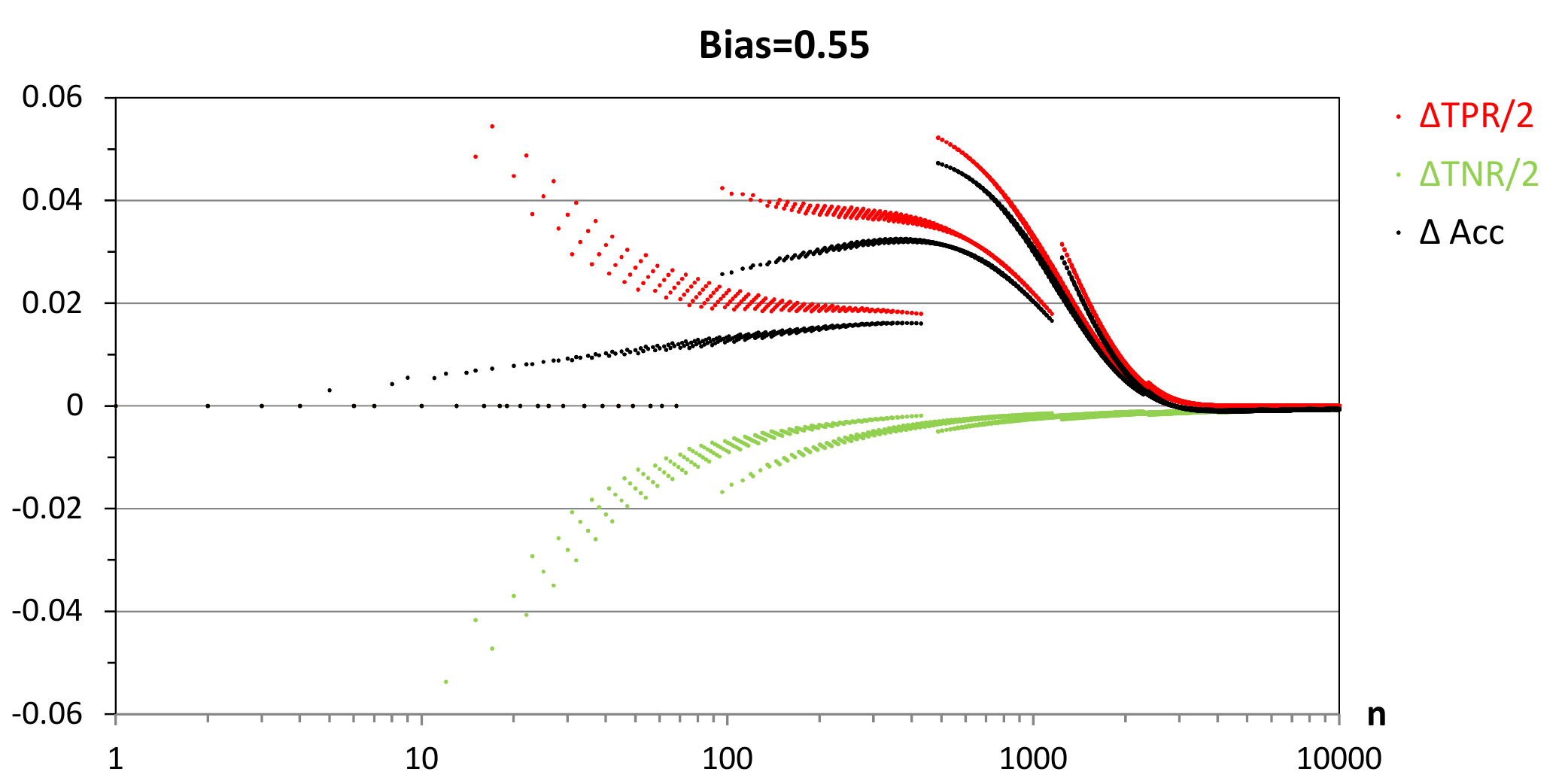}\\
\includegraphics[width=0.45\textwidth]{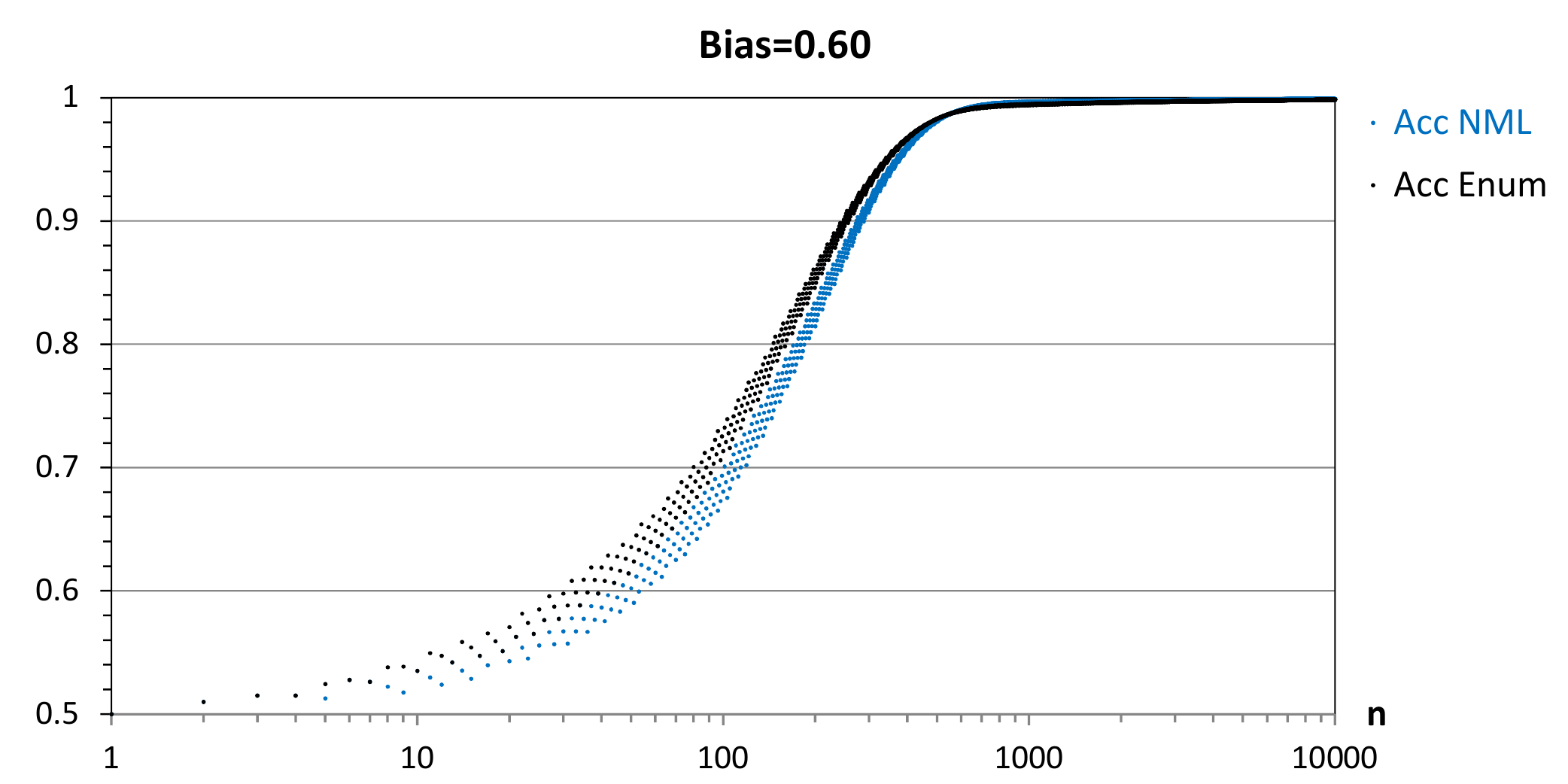}
\includegraphics[width=0.45\textwidth]{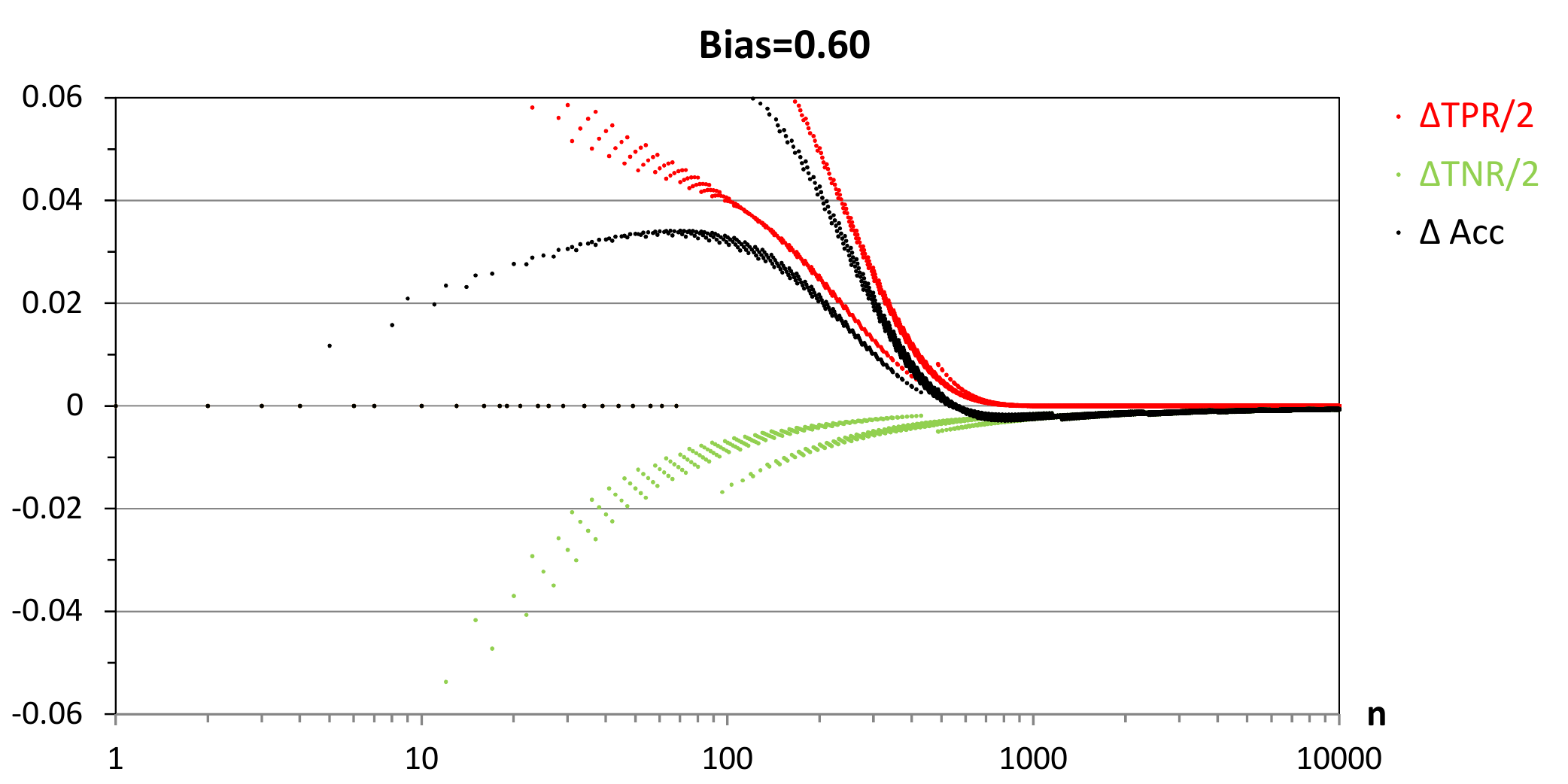}\\
\includegraphics[width=0.45\textwidth]{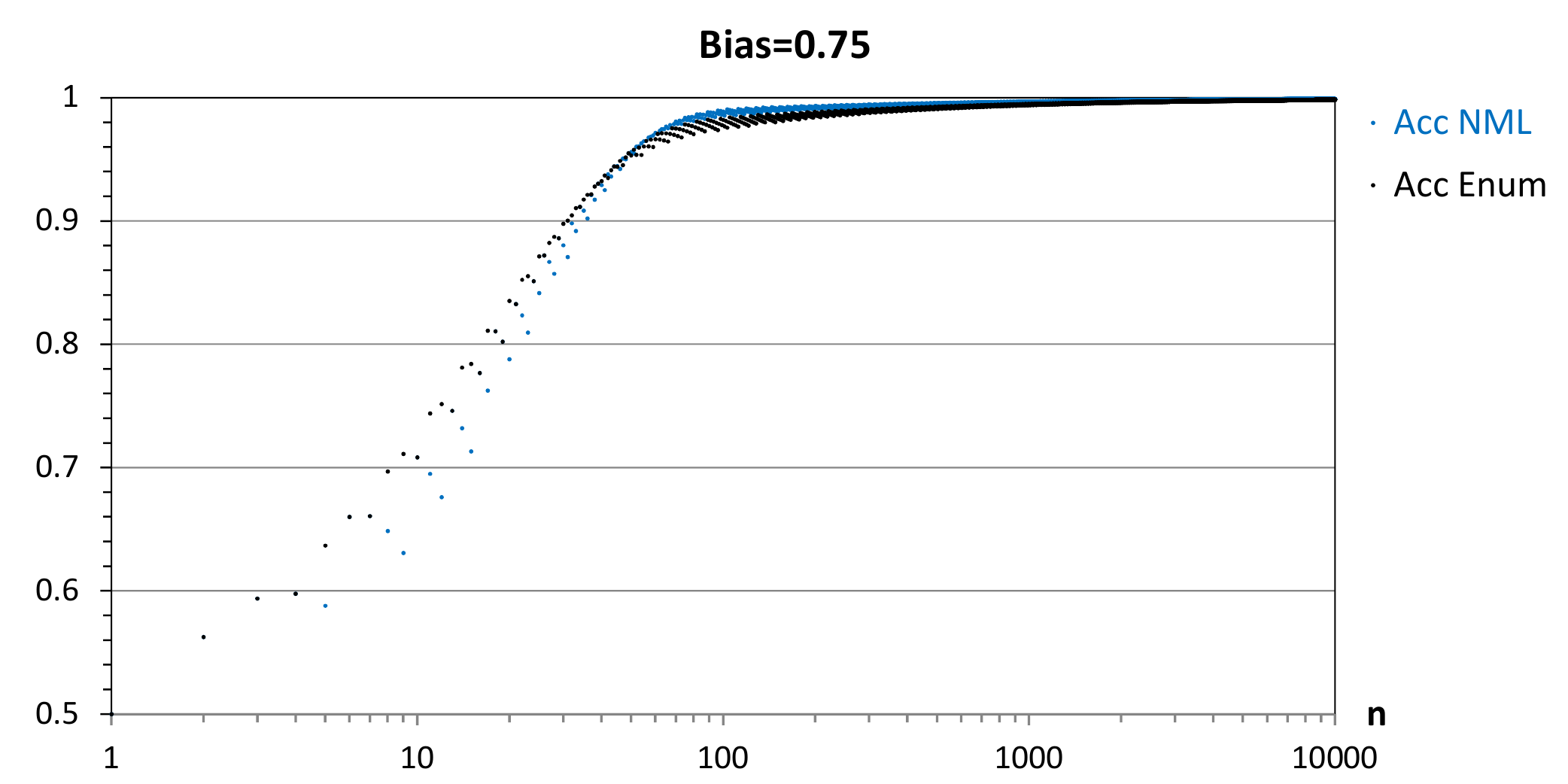}
\includegraphics[width=0.45\textwidth]{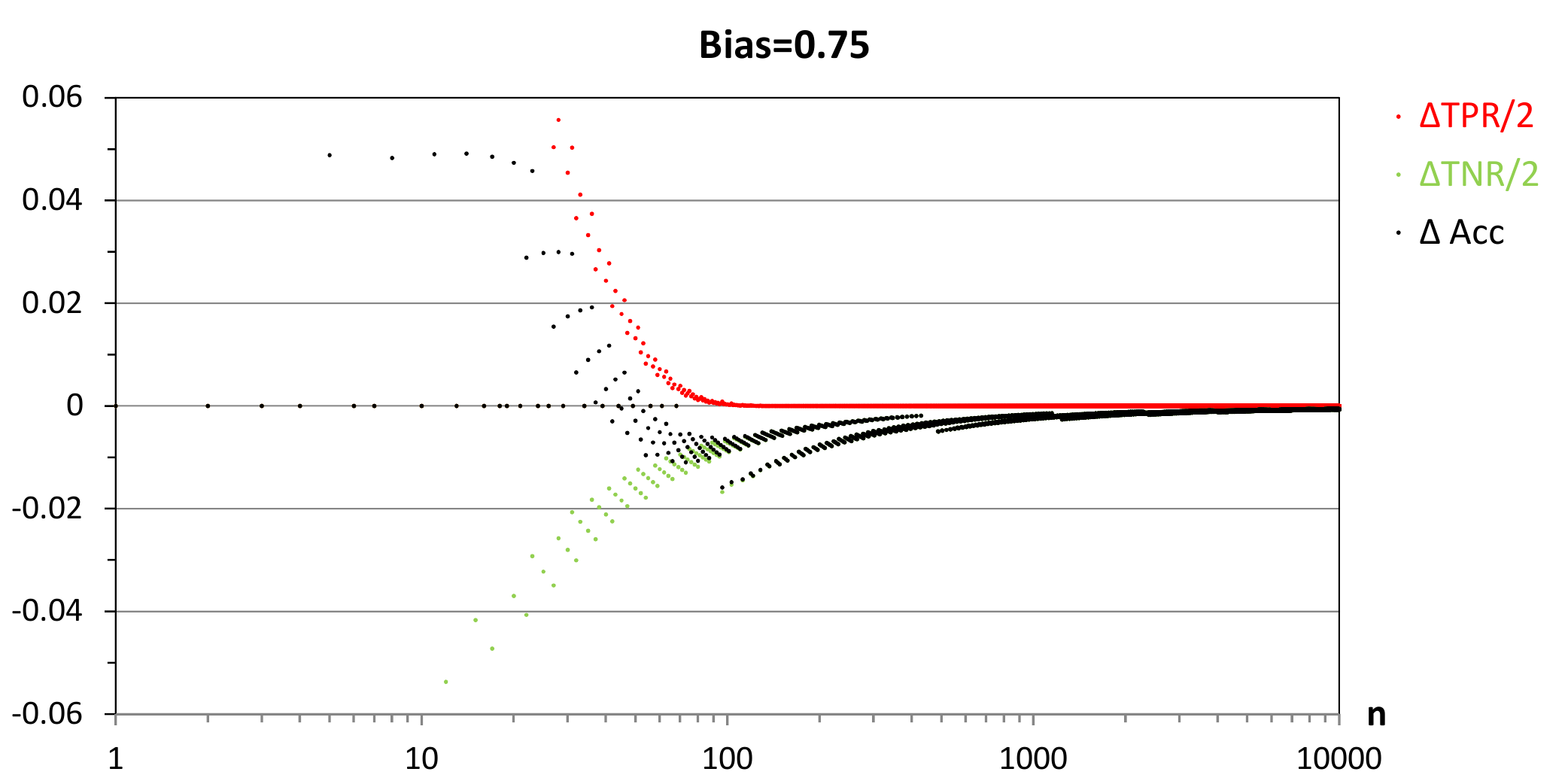}\\
\includegraphics[width=0.45\textwidth]{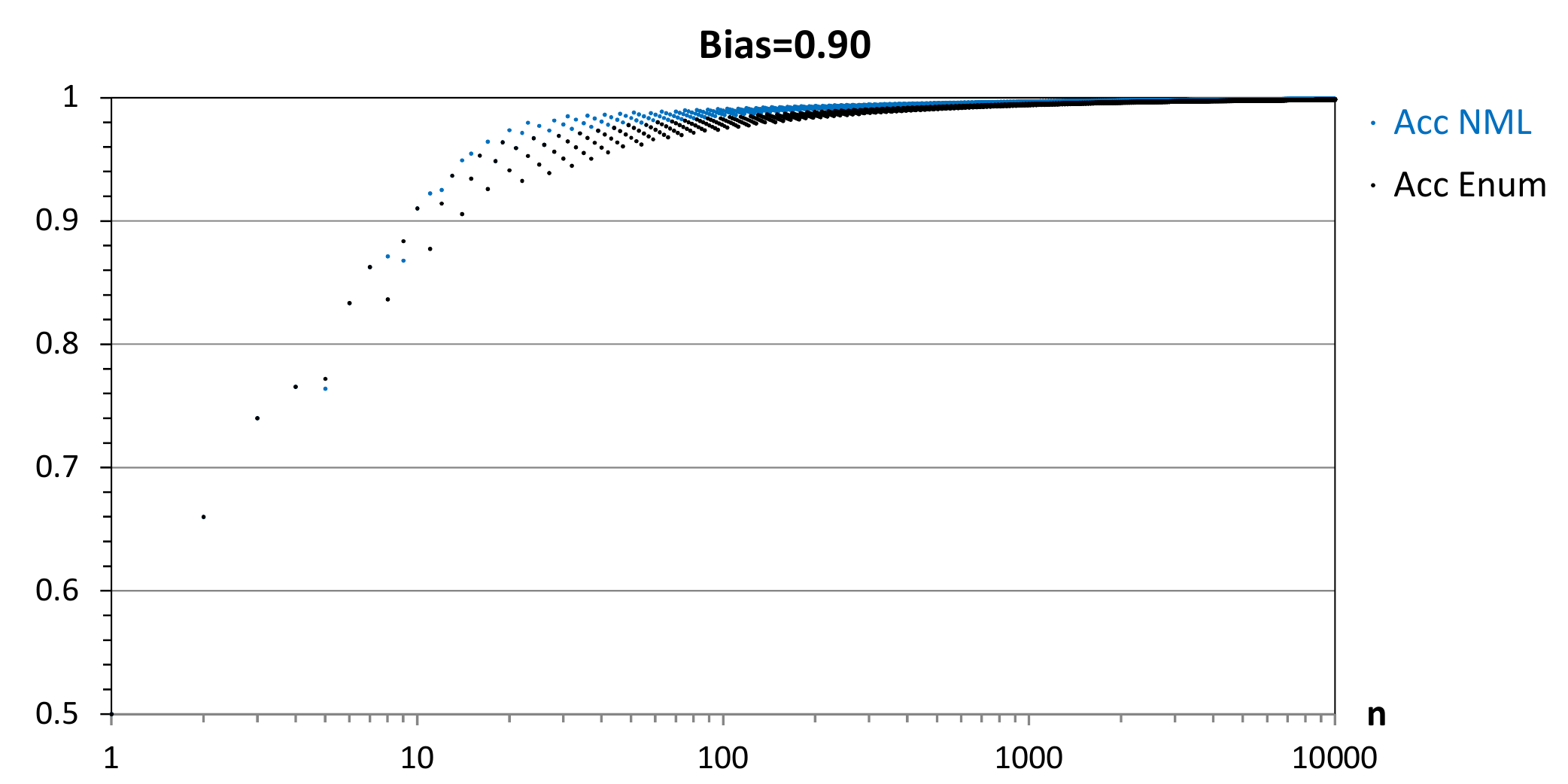}
\includegraphics[width=0.45\textwidth]{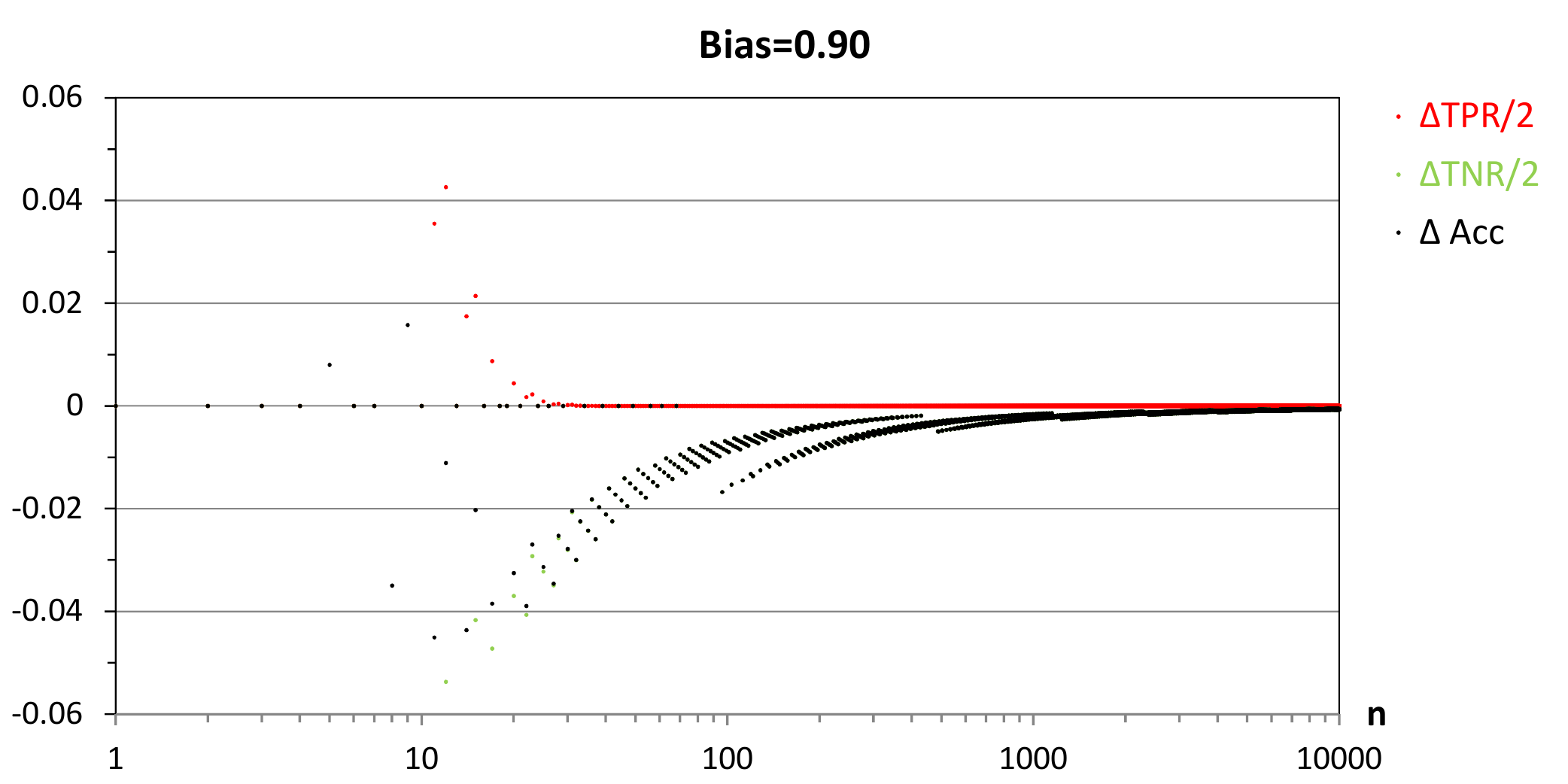}
 \caption{Classification of coins using the NML and Enum codes for different biases.
Accuracy (left) and difference $(Enum - NML)$ for the true positive, false negative and accuracy (right).}
 \label{fig_CoinClassification}
\end{center}
\end{figure}

We perform the coin classification experiment for $\theta_{bias} \in \{0.501, 0.51, 0.55, 0.75, 0.60, 0.90\}$ and $n$ ranging from 1 to $10,000$ using the enumerative and the NML codes ($n$ up to $10,000,000$ for $\theta_{bias}  = 0.501$).
Figure~\ref{fig_CoinClassification} reports the accuracy results (left) as well as the difference $(Enum - NML)$ of the three indicators.

Overall, both codes exhibit a similar behavior w.r.t. the coin classification problem, with accuracy increasing from 0.5 for small $n$ to 1 for large $n$, and a slow increase rate for small bias and a fast one for large bias. Except in the tiny samples with $n \leq 20$, the difference between any of the three indicators never exceeds around 15\%.
However, there are some interesting differences.
As noticed in Section~\ref{biasedCoin}, the enumerative code has a better sensitivity at the expense to a worse specificity, and the aggregated accuracy result exhibits a variety of behaviors.
When the bias is small ($\theta_{bias}$ close from $\frac{1}{2}$), the enumerative code is far more sensitive while being a little less specific, resulting in more accurate predictions in the non-asymptotic case.
When the bias is large ($\theta_{bias}$ far from $\frac{1}{2}$), both codes get almost the same sensitivity while the enumerative code remains less specific, resulting in slightly less accurate predictions.
In all cases, the differences between both codes get tiny for large $n$, in the asymptotic case.

\section{The case of multinomial distribution}
\label{secMultinomial}

Let us consider the multinomial model with parameter $\theta = (\theta_1, \ldots, \theta_m), \; \sum_{j=1}^m {\theta_j}=1, \forall j, \theta_j> 0$, such that $P_{\theta}(X=j) = \theta_j$, in the case of m-ary sequences $x^n \in X^n$ of size $n$.
For a given sequence $x_n$, $P_{\theta}(x_n) = \prod_{j=1}^m {\theta_j^{n_j}}$, where $n_j$ is the number of occurrences of outcome $j$ in sequence $x^n$.

\subsection{Standard NML approach}
\label{secMultinomialNML}

 As pointed out in \citep{RooijEtAl09},
\begin{quotation}
``The NML distribution has a number of significant practical problems.
First, it is often undefined, because for many models the denominator in (\ref{eqnNML}) is infinite, even for such simple models as the Poisson or geometric distributions. 
Second, $X^n$ is exponentially large in $n$, so calculating the NML probability exactly is only possible in special cases such as the Bernoulli model above, where the number of terms is reduced using some trick. Something similar is possible for the more general multinomial model 
(...)
, but in most cases 
[it] 
has to be approximated, which introduces errors that are hard to quantify.''
\end{quotation}

The parametric complexity of the NML universal model with respect to a k-parameter exponential family model is usually approximated by $\frac{k}{2} \log \frac{n}{2 \pi}$ \citep{Grunwald07}.
In the case of the multinomial distribution with $(m-1)$ free parameters, this gives $\frac{m-1}{2} \log \frac{n}{2 \pi}$.
A better approximation based on Rissanen's asymptotic expansion \citep{Rissanen96} is presented in \citep{Kontkanen2009}:
\begin{equation}
\label{compM_R}
COMP_{nml}^{(n)}(\theta) = \frac{m-1}{2} \log \frac{n}{2 \pi} + \log \frac{\pi^{m/2}}{\Gamma (m/2)} + o(1),
\end{equation}
where $\Gamma(.)$ is the Euler gamma function.
Still in \citep{Kontkanen2009}, a sharper approximation based on Szpankowski's approximation is presented. This last approximation, far more complex is very accurate w.r.t. $n$, with $o(\frac{1}{n^{3/2}})$ precision.
We present below its first terms until $o(\frac{1}{\sqrt n})$, which actually are the same that in Rissanen's approximation:
\begin{equation}
\label{compM_S}
COMP_{nml}^{(n)}(\theta) = \frac{m-1}{2} \log \frac{n}{2} + \log \frac{\sqrt \pi}{\Gamma (m/2)} + o(\frac{1}{\sqrt n}),
\end{equation}

Finally, \citep{KontkanenEtAl07} propose an exact computation of the multinomial stochastic complexity, at the expense of sophisticated algorithms with quasilinear computation time.

\subsection{Enumerative two-part crude MDL}
\label{secEnumerativeM}

We apply the same approach as in the case of the Bernoulli model (Sections~\ref{secEnumerativeB} and \ref{secEnumBayesian}).
Given a sample size $n$, the number of tuples $(n_1, n_2, \ldots, n_m)$ such that $\sum_{j=1}^m {n_j} = n$ is ${{n+m-1}\choose{m-1}}$. 
We then encode the multinomial model parameter using a uniform prior 
$$P\left(\theta = \left(\frac{n_1}{n}, \frac{n_2}{n}, \ldots, \frac{n_m}{n}\right) \right) = 1 / {{n+m-1}\choose{m-1}},$$ leading to $L(\theta) = \log {{n+m-1}\choose{m-1}}$.
Second, we have to encode the data $x^n$ at best given the $\theta$ parameter.

We suggest using a probability distribution for encoding the finite size data sample $x^n$, with the following likelihood.

For $\theta \neq \left(\frac{n_1(x^n)}{n}, \frac{n_2(x^n)}{n}, \ldots, \frac{n_m(x^n)}{n}\right)$, we cannot encode the data and $P(x^n|\theta) = 0$.

For $\theta = \widehat{\theta}(x^n) = \left(\frac{n_1(x^n)}{n}, \frac{n_2(x^n)}{n}, \ldots, \frac{n_m(x^n)}{n}\right)$, the observed data is consistent with the model parameter and we assume that all the possible observable data are uniformly distributed.
The number of m-ary strings where the number of occurrences of outcome $j$ is $n_j$ is given by the multinomial coefficient $\frac{n!}{n_1! n_2! \ldots n_m!}$. Thus the probability of observing one particular m-ary string is $P(x^n|\widehat{\theta}(x^n)) = 1/\frac{n!}{n_1! n_2! \ldots n_m!}$.
This gives a total code length of

\begin{equation}
L(\widehat{\theta}(x^n), x^n) = \log {{n+m-1}\choose{m-1}} + \log \frac{n!}{n_1! n_2! \ldots n_m!},
\end{equation}
defined only when $\theta = \widehat{\theta}(x^n)$.

\subsection{NML interpretation}

Let us compute the NML parametric complexity of this enumerative code on the basis of the discrete  likelihood.
We have

\begin{eqnarray}
COMP^{(n)}(\theta) &=& \log \sum_{y^n \in X^n} {P_{\widehat{\theta}(y^n)}(y^n)}, \\
  &=& \log \sum_{\{n_1+\ldots+n_m=n\}} {\frac{n!}{n_1! n_2! \ldots n_m!} \left({1} / \frac{n!}{n_1! n_2! \ldots n_m!} \right)},\\
	&=& \log {{n+m-1}\choose{m-1}}.
\end{eqnarray}

Interestingly, we find exactly the same complexity term as the coding length of the best hypothesis in the enumerative approach, that simply relies on counting the possibilities for the model parameters.
Like in the Bernoulli case, this shows that the enumerative code is both a two-part and a one-part code, optimal w.r.t. the NML approach and parametrization invariant.
We have an exact formula for the complexity term, very simple to compute.
Using Stirling's approximation $\log n! = n \log n - n + \frac{1}{2} \log {2 \pi n} + O(1/n)$, we get the following asymptotic approximation:

\begin{equation}
\label{compEnum_M}
COMP^{(n)}(\theta) = (m-1)(\log n - \log (m -1) +1) -\frac{1}{2} \log {2 \pi (m-1)} + o(\frac{1}{n}).
\end{equation}

Once again, this asymptotic model complexity is twice that of the alternative classical NML code or the standard BIC regularization term $(m-1)\log n$.

\section{Code comparison for the multinomial distributions}
\label{secComparisonM}

In this section, we compare the NML code (Section~\ref{secMultinomialNML}) and enumerative two-part crude MDL codes (Section~\ref{secEnumerativeM}) for the multinomial distribution.

\subsection{Notation}
\label{secNotationM}

Let us use the names \emph{NML} and \emph{enumerative} for the specific MDL codes presented in Sections \ref{secMultinomialNML} and \ref{secEnumerativeM}.
We also consider the \emph{random} code as a baseline: it corresponds to a direct encoding of each binary string $x^n$ with a coding length of $n \log m$. The likelihood of each string $x^n$ is $1/m^n$, and as $\sum_{\{n_1+\ldots+n_m=n\}} {\frac{n!}{n_1! n_2! \ldots n_m!}  1/m^n} = 1$, we have $COMP_{random}^{(n)}(\emptyset) = 0$ and 
$L_{random}\left(x^n|\emptyset \right) = n \log m$.

Table~\ref{tableCodesM} reminds the parametric and stochastic complexity of each considered code for the multinomial distribution.

\begin{table}[htbp!]
\caption{Parametric and stochastic complexity per code.}
\label{tableCodesM}
\centering
\renewcommand{\arraystretch}{1.5}
\begin{tabular}[10pt]{ccc}\hline
Code name         & $COMP_{name}^{(n)}$ & $L_{name}\left(x^n|\widehat{\theta}(x^n) \right)$ \\\hline 
\emph{enumerative}      & $\log {{n+m-1}\choose{m-1}}$    & $\log \frac {n!} {n_1! \ldots n_m!}$ \\
\emph{NML}  & $\frac{m-1}{2} \log \frac{n}{2} + \log \frac{\sqrt \pi}{\Gamma (m/2)} + o(\frac{1}{\sqrt n})$ & $\log \frac {n^n} {n_1^{n_1} \ldots n_m^{n_m}}$ \\
\emph{random}     & $0$    & $n \log m$ \\\hline
\end{tabular}
\end{table}

\subsection{Stochastic complexity term}
\label{secStochasticM}

The stochastic complexity term of the enumerative code is always smaller than that of the NML code for non-degenerated m-ary strings:

\begin{equation}
\label{sc_Multinomial}
	\forall n, \forall x^n \in X^n \; \mbox{such that} \; (\max_{1 \leq j \leq m} n_j ) < n \; \mbox{ then} \;
	L_{enum}\left(x^n|\widehat{\theta}(x^n)\right) < L_{nml}\left(x^n|\widehat{\theta}(x^n)\right).
\end{equation}

An intuitive proof relies on the fact that the enumerative MDL likelihood assigns the same probability to all m-ary strings having the same number of occurrence per outcome $j$, with a null probability for all the other strings. The NML likelihood also assigns the same probability to these m-ary strings, but with a non-null probability for the other strings. Then they have to share a smaller probability mass, resulting in a smaller probability per string and a strictly greater coding length.

\medskip

To gain further insights, let us approximate the difference of coding length for the stochastic complexity term:
$$\delta L_{SC}\left(x^n|\widehat{\theta}(x^n)\right) = L_{nml}\left(x^n|\widehat{\theta}(x^n)\right) -L_{enum}\left(x^n|\widehat{\theta}(x^n)\right).$$

We assume that $\forall j, n_j > 0$ and $n_j \approx n \widehat{\theta}_j$.
Using Stirling's approximation $\log n! = n \log n - n + \frac{1}{2} \log {2 \pi n} + O(1/n)$, we get
\begin{eqnarray*}
L_{enum}\left(x^n|\widehat{\theta}(x^n)\right)
  &=& \log \frac{n!}{n_1! n_2! \ldots n_m!} \\
  &=& n \log n -n + \frac{1}{2} \log {2 \pi n} + O(1/n) \\ 
  && -\sum_{j=1}^m {\left(n_j \log n_j - n_j + \frac{1}{2} \log {2 \pi n_j} + O(1/n_j)\right)} \\
	&=& \log \frac {n^n} {n_1^{n_1} \ldots n_m^{n_m}} - \frac{m-1}{2} \log {2 \pi n} 
		- \frac{1}{2} \log \prod_{j=1}^m {\widehat{\theta}_j} + O(1/n) \\
\end{eqnarray*}

We get

\begin{equation}
\label{deltaM_SC}
\delta L_{SC}\left(x^n|\widehat{\theta}(x^n)\right) = \frac{m-1}{2} \log {2 \pi n} 
+ \frac{1}{2} \log \prod_{j=1}^m {\widehat{\theta}_j} + O(1/n).
\end{equation}

It is noteworthy that the $\mathrm{var}(\widehat{\theta})$ term in the Bernoulli case (see Formula~\ref{deltaL}) generalizes to a $\prod_{j=1}^m {\widehat{\theta}_j}$ term in the multinomial case.

\medskip
These results demonstrate that the enumerative code provides a better encoding of the data with the help of the model for any m-ary strings. The gain in coding length compared to the NML code is always positive and grows asymptotically as $(m-1)/2$ times the logarithm of the sample size.

\subsection{Parametric complexity term}
\label{secParametricM}

Using inequality~\ref{sc_Multinomial} and as the parametric complexity of code is the sum of the stochastic complexity over all possible strings, we get:

\begin{equation}
\label{pc_Multinomial}
	\forall n > 1, COMP_{enum}^{(n)} > COMP_{NML}^{(n)}.
\end{equation}

Both terms are equal for $n=1$ and asymptotically, the parametric complexity of the enumerative code is twice that of the NML code (see Formulas~\ref{compM_S} and \ref{compEnum_M}).

\begin{figure}[!htb]
\begin{center}
 \includegraphics[width=0.85\textwidth]{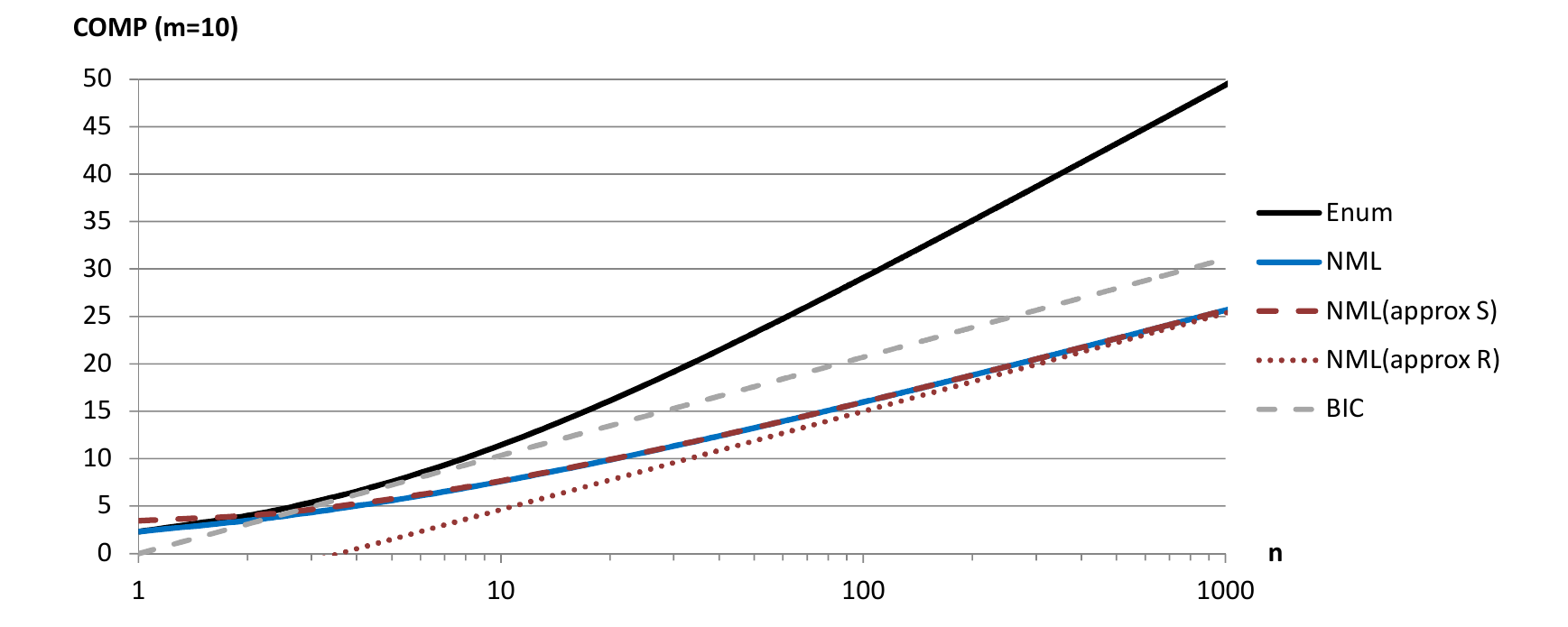}\\
 \includegraphics[width=0.85\textwidth]{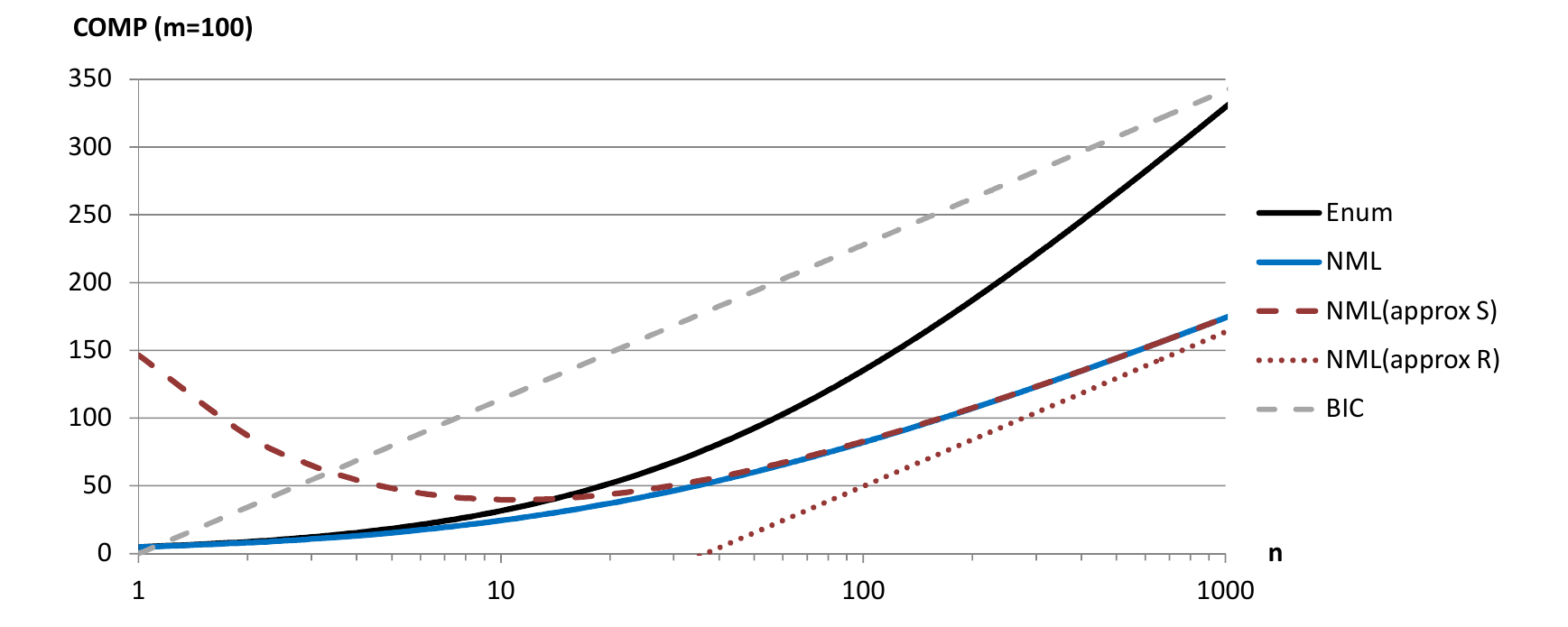}
 \caption{Parametric complexity for the multinomial model with 10 or 100 categories.}
 \label{fig_compM}
\end{center}
\end{figure}

We now focus on the non-asymptotic behavior of the parametric complexity terms and their approximations.
Figure~\ref{fig_compM} shows the value of the parametric complexity of the multinomial model, using the enumerative code, the NML code (exact numerical computation and Rissanen or Szpankowski based approximations: see Section~\ref{secMultinomialNML}), as well as the related BIC penalization term.

The BIC approximation is very bad, all the more as $m$ increases.
The Rissanen approximation of the NML parametric complexity is very good as soon as $n$ is about 100 times the number $m$ of categories, but less accurate for small sample sizes.
As expected, the Szpankowski based approximation if much sharper, being accurate as soon as $n$ is beyond $m$, but with bad accuracy for $n << m$.

\medskip
Let us now compute an asymptotic approximation of the difference of parametric complexity between the two codes:
$$\delta L_{PC} COMP^{(n)}(\theta) = COMP_{nml}^{(n)}(\theta) - COMP_{enum}^{(n)}(\theta).$$

Using previous approximations presented in Formulas~\ref{compM_S}, ~\ref{compEnum_M} and the Stirling's approximation of the Gamma function, we get:

\begin{eqnarray}
\label{eqn_diffPC_M}
\delta L_{PC} COMP^{(n)}(\theta)
 &=& \frac{m-1}{2} \log \frac{n}{2} + \log \sqrt \pi 
-(\frac{m}{2} \log \frac{m}{2} - \frac{m}{2} -\frac{1}{2} \log \frac{m}{4 \pi})\\
&& - ((m-1)(\log n - \log (m -1) +1) -\frac{1}{2} \log {2 \pi (m-1)}) \\
&& + o(\frac{1}{\sqrt n}),\\
 &=& -\frac{m-1}{2} \log n -\frac{m-1}{2} \log 2 + \frac{1}{2}\log \pi \\
&&-\frac{m}{2} \log m + \frac{m}{2} \log 2 + \frac{m}{2} + \frac{1}{2} \log m - \frac{1}{2} \log \pi - \log 2\\
&& + (m-1)\log (m -1) - (m-1) \\
&& + \frac{1}{2} \log 2 + \frac{1}{2} \log \pi + \frac{1}{2} \log (m-1) + o(\frac{1}{\sqrt n}),\\
&=& -\frac{m-1}{2} \log {n m e} + (m - \frac{1}{2}) \log (m-1) + \frac{1}{2} \log {\pi e} + o(\frac{1}{\sqrt n}).
\end{eqnarray}

We obtain
\begin{eqnarray}
\label{deltaM_PC}
\delta L_{PC} COMP^{(n)}(\theta)
 &=& -\frac{m-1}{2} \log n + \frac{m}{2}\log{\frac{m}{e}} \\ \nonumber
 && +\log \frac{e}{\sqrt \pi} + (m-\frac{1}{2})\log(1-\frac{1}{m}) + o(\frac{1}{\sqrt n})
\end{eqnarray}

 and for $n >> m$

\begin{eqnarray}
\label{deltaM_PC2}
\delta L_{PC} COMP^{(n)}(\theta)
 &=& -\frac{m-1}{2} \log n + \frac{m}{2}\log{\frac{m}{e}} 
 -\log \sqrt \pi + o(\frac{1}{m}) + o(\frac{1}{\sqrt n}).
\end{eqnarray}

This result demonstrates that the difference of parametric complexity increases as the logarithm of the sample size. The speed of increase is with a factor $(m-1)/2$, but for small sample sizes (typically $n \leq m$), the difference remains small.

\begin{figure}[!htb]
\begin{center}
 \includegraphics[width=0.85\textwidth]{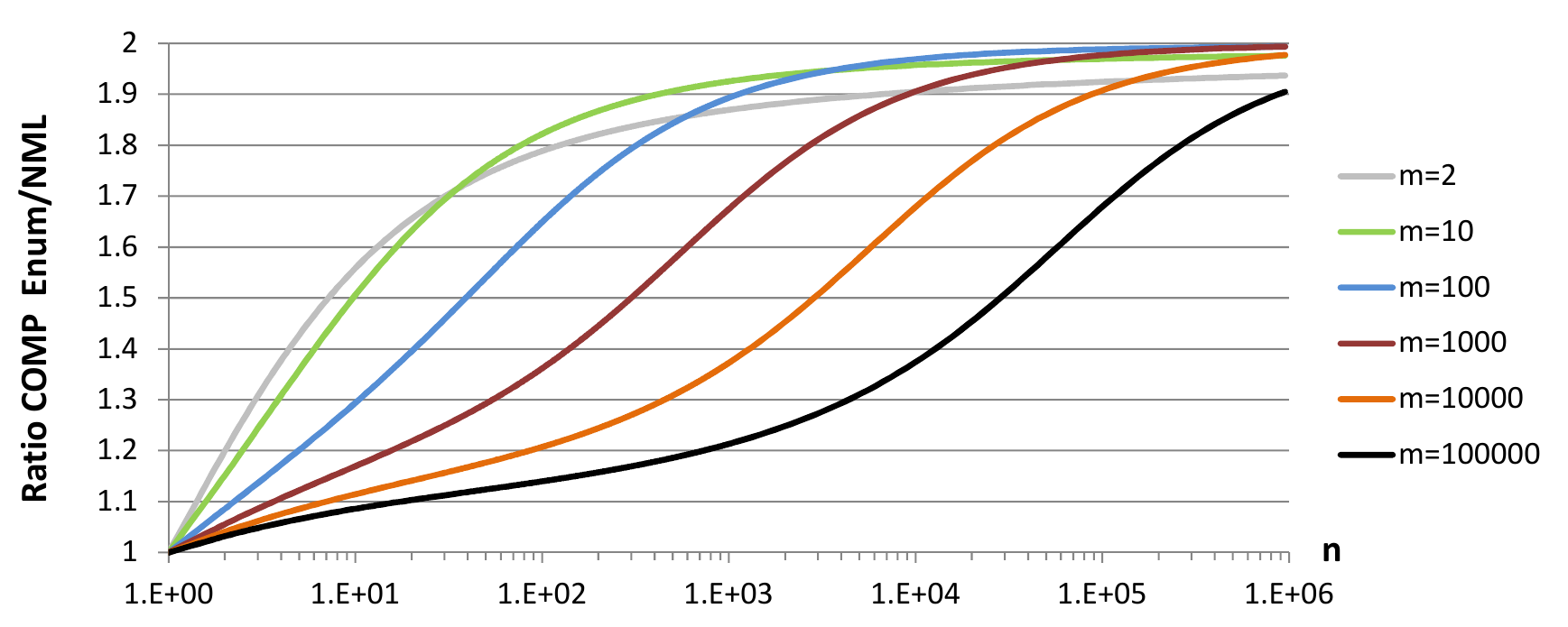}
 \caption{Ratio of enumerative to the NML parametric complexities for the multinomial model with up to 100,000 outcomes.}
 \label{fig_ratio_compM}
\end{center}
\end{figure}

We illustrate this behavior in the non-asymptotic case.
Figure~\ref{fig_ratio_compM} focuses  on the ratio of the exact parametric complexity terms for the enumerative and NML codes. This ratio always increases from 1 for $n=1$ to 2 when $n$ goes to infinity, with the speed of convergence decreasing as the number $m$ of outcomes increases.

\subsection{Overall code length}
\label{secOverallM}

Both the enumerative and NML codes exploit universal distributions on all m-ary strings $x^n \in X^n$.
The compression of the data with the help of the model is better for the enumerative distribution, at the expense of a worse parametric complexity.
The overall code length is the sum of the parametric and stochastic complexities.
Using previous approximations presented in Formulas~\ref{deltaM_SC}, ~\ref{deltaM_PC}, we obtain the following approximation of the difference of overall code lengths between the two codes:

\begin{eqnarray}
\label{eqn_mixtureM}
\Delta L\left(\widehat{\theta}(x^n), x^n\right)
&=& \delta L_{PC} COMP^{(n)}(\theta) + \delta L_{SC}\left(x^n|\widehat{\theta}(x^n)\right),\\
&=& -\frac{m-1}{2} \log n + \frac{m}{2}\log{\frac{m}{e}}
 +\log \frac{e}{\sqrt \pi} + (m-\frac{1}{2})\log(1-\frac{1}{m})\\
&& + \frac{m-1}{2} \log {2 \pi n} 
+ \frac{1}{2} \log \prod_{j=1}^m {\widehat{\theta}_j} + o(\frac{1}{\sqrt n}).
\end{eqnarray}

We obtain
\begin{footnotesize}
\begin{eqnarray}
\label{eqn_mixtureM2}
\Delta L\left(\widehat{\theta}(x^n), x^n\right)
 &=& \frac{m}{2} \log \frac{m 2 \pi}{e} + \frac{1}{2} \log \prod_{j=1}^m {\theta_j}
 +\log \frac{e}{\sqrt 2} + (m-\frac{1}{2})\log(1-\frac{1}{m}) 
 + o(\frac{1}{\sqrt n}),
\end{eqnarray}
\end{footnotesize}

 and for $n >> m$

\begin{eqnarray}
\label{eqn_mixtureM3}
\Delta L\left(\widehat{\theta}(x^n), x^n\right)
 &=& \frac{m}{2} \log \frac{m 2 \pi}{e}
 + \frac{1}{2} \log \prod_{j=1}^m {\theta_j} 
 - \log \sqrt 2 + o(\frac{1}{m}) + o(\frac{1}{\sqrt n}) .
\end{eqnarray}

Asymptotically, the difference in overall code length does not depend on the size $n$ of the string.
Both codes differ by a margin that depends essentially on the number $m$ of outcomes and of the multinomial parameter $\theta$.

\paragraph{Case of balanced multinomial distributions.}
The term $\log \prod_{j=1}^m {\theta_j}$ is minimal for equidistributed multinomial distribution ($\theta_j=1/m$).
For such distributions, we get 
\begin{eqnarray}
\label{eqn_mixtureM4}
\Delta L\left(\widehat{\theta}(x^n), x^n\right)
 &=& \frac{m}{2}\log \frac{2 \pi}{e} + \log \frac{e}{\sqrt 2} + o(\frac{1}{m}) + o(\frac{1}{\sqrt n}),
\end{eqnarray}
which means that the enumerative code compresses the strings better than the NML code with a margin that increases linearly with $m$.

\paragraph{Case of degenerated multinomial distributions.}
In case of multinomial distributions with one single observed outcome (e.g. $\widehat{\theta} = (1, 0, \ldots, 0)$), both the NML and enumerative codes have a null stochastic complexity and the difference between the coding lengths reduces to the difference between the parametric complexity terms (see Formula~\ref{deltaM_PC}). In this extreme case, the NML code compresses the string better with a margin that grows as $\frac{m-1}{2}$ times the logarithm of the sample size.

\paragraph{Case of unbalanced multinomial distributions.}
Let us study the boundary between balanced distributions and degenerated distributions, where the enumerative code dominates the NML code and conversely.
We are seeking for distributions where both codes achieve approximately the same compression. 
For that purpose, let us consider peaked multinomial distributions, with most of the probability mass for the first outcome ($\theta_1 = \theta_{max}$) and the rest of the probability mass equistributed for the other outcomes ($\theta_j = \theta_{min}, 2 \leq j \leq m$, with $\theta_{min} = \frac {1 - \theta_{max}} {m-1}$).
Using Formula~\ref{eqn_mixtureM3}, we thus try to find the peaked distribution $\theta = (\theta_{max}, \theta_{min}, \ldots, \theta_{min})$ such that $\Delta L\left(\widehat{\theta}(x^n), x^n\right) = o(1)$.
The solution is obtained for

\begin{equation}
\label{peakThresholdMax}
\theta_{max} = 1-\frac{m-1}{m + \log m}\frac{e}{2 \pi} > 0.56,
\end{equation}

\begin{equation}
\label{peakThresholdMin}
\theta_{min} = \frac{e}{2 \pi (m + \log m)} < 0.44\frac{1}{m},
\end{equation}

leading to

\begin{equation}
\Delta L\left(\widehat{\theta}(x^n), x^n\right) = \alpha + o(\frac{1}{m}) + o(\frac{1}{\sqrt n})
\end{equation}

with $\alpha \approx 0.37$. This peaked distribution is at the limit where the NML code compresses the data better than the enumerative code.
Numerical experiments, not reported here, confirm the accuracy of Formulas~\ref{peakThresholdMax} and \ref{peakThresholdMin} and show that the asymptotic value of the peak probability $\theta_{max}$ behaves as a lower bound of the obtained probability in the non asymptotic case.

For Bernoulli distributions, we had $\theta_{max} \approx 0.886$ (see Formula~\ref{eqn_compareCodes}), and not surprisingly, $\theta_{max}$ decreases with $m$ (see Formula~\ref{peakThresholdMax}).
Interestingly, $\theta_{max}$ is always greater than $0.56$ whatever $m$. This means that when $m$ increases, the ratio $\theta_{max}/\theta_{min}$ grows linearly with $m$ and the fraction of multinomial distributions where the NML code dominates the enumerative code becomes negligible.

\paragraph{Synthesis.}

The overall difference of coding length is in favor of the enumerative code for balanced strings with a margin that increases linearly with $m$. The NML code is better only for heavily unbalanced strings, where the most frequent outcome occurs more that half of the times, whatever be $m$.
Under the uniform distribution, such unbalanced strings are far less frequent than balanced strings, and the enumerative code compresses most strings better than the NML code.

\subsection{Expectation of the coding length of all m-ary strings}

\begin{figure}[!htb]
\begin{center}
 \includegraphics[width=0.49\textwidth]{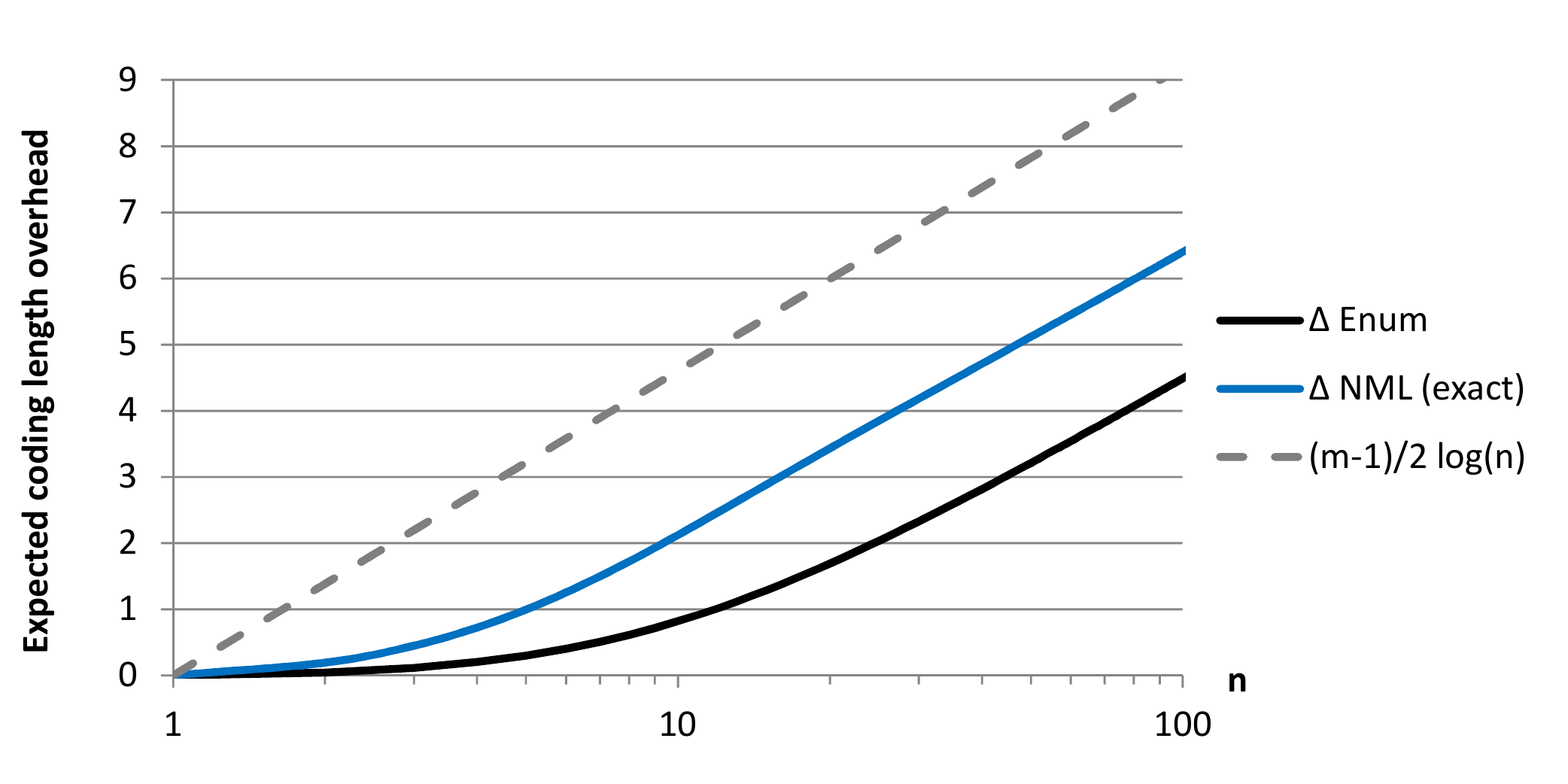}
 \includegraphics[width=0.49\textwidth]{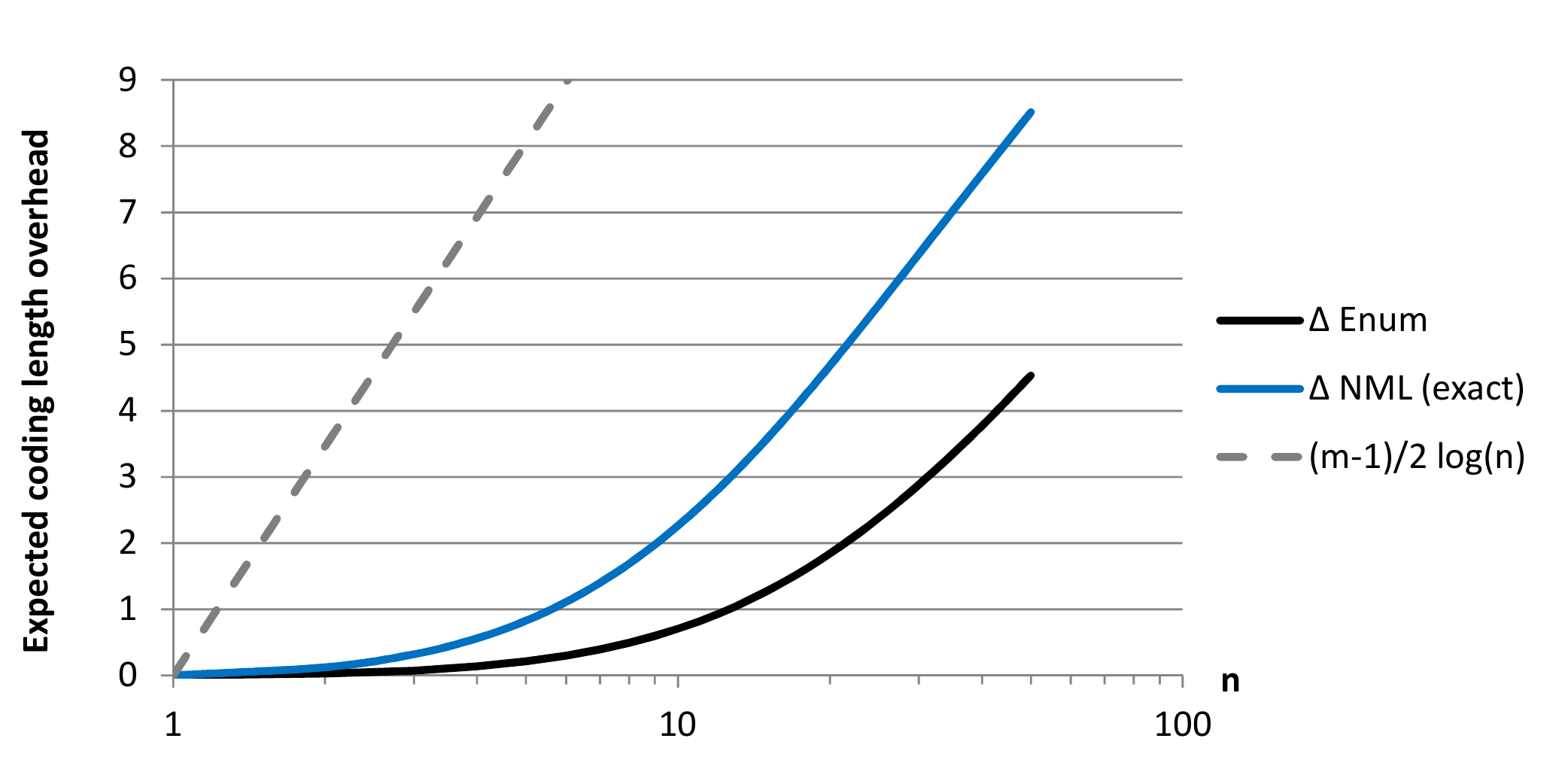}
 \caption{Expected overhead of coding length w.r.t. random model for $m=5$ (left) and $m=10$ (right).}
 \label{fig_codinglengthM}
\end{center}
\end{figure}

Let us estimate the expectation of the coding length for all m-ary strings under the uniform distribution, where $\forall x^n \in X^n, p(x^n) = 1/{m^n}$.
\begin{eqnarray}
\label{eqn_codinglengthM}
\mathrm{E}\left(L\left(\widehat{\theta}(x^n), x^n\right)\right) &=& \frac{1} {m^n} \sum_{x^n \in X^n} {L\left(\widehat{\theta}(x^n), x^n\right)},\\ \nonumber
  &=& \frac{1} {m^n} \sum_{\{n_1+\ldots+n_m=n\}} {\frac{n!}{n_1! n_2! \ldots n_m!} L\left(\widehat{\theta}(x^n), x^n\right)}.
\end{eqnarray}

We perform an exact numerical calculation using the exact value of the NML parametric complexity term, for $m=5$ ($1 \leq n \leq 100$) and $m=10$ ($1 \leq n \leq 50$). Let us notice that the sum in Formula~\ref{eqn_codinglengthM} involves more than ten billions terms for $m=10$ and $n=50$.
Figure~\ref{fig_codinglengthM} reports the expected coding length for the enumerative and NML codes minus that of the random code ($n \log m$).
The results show that both codes have an asymptotic overhead that grows as $(m-1)/2 \log n$, compared to the direct encoding of the binary strings. 
Under the uniform distribution, the enumerative code always compresses the data better than the NML code, especially in the non-asymptotic case. 
As for the Bernoulli codes, most possible m-ary strings are almost equidistributed and their shorter coding lengths obtained with the enumerative provide the main contribution in the expectation of the coding length.
Following Formula~\ref{eqn_mixtureM3}, the enumerative code compresses the m-ary strings better than the NML code with a margin that grows linearly with $m$.

\subsection{Percentage of compressible m-ary strings}
\label{percentCompressibleMultibomial}

We now focus on the percentage $p_{compressible}$ of compressible m-ary strings using both the enumerative and NML codes, that is the percentage of m-ary strings with coding length shorter than $n \log m$:

\begin{eqnarray}
\label{eqn_percentMcompressible}
p_{compressible} &=& \frac{1} {m^n} \sum_{x^n \in X^n}
 {\mathbb{1}_{\left\{L\left(\widehat{\theta}(x^n), x^n\right) \leq n \log m\right\}}},\\ \nonumber
  &=& \frac{1} {m^n} \sum_{\{n_1+\ldots+n_m=n\}} {\frac{n!}{n_1! n_2! \ldots n_m!} 
	\mathbb{1}_{\left\{L\left(\widehat{\theta}(x^n), x^n\right) \leq n \log m\right\}}}.
\end{eqnarray}

\begin{figure}[!htb]
\begin{center}
 \includegraphics[width=0.49\textwidth]{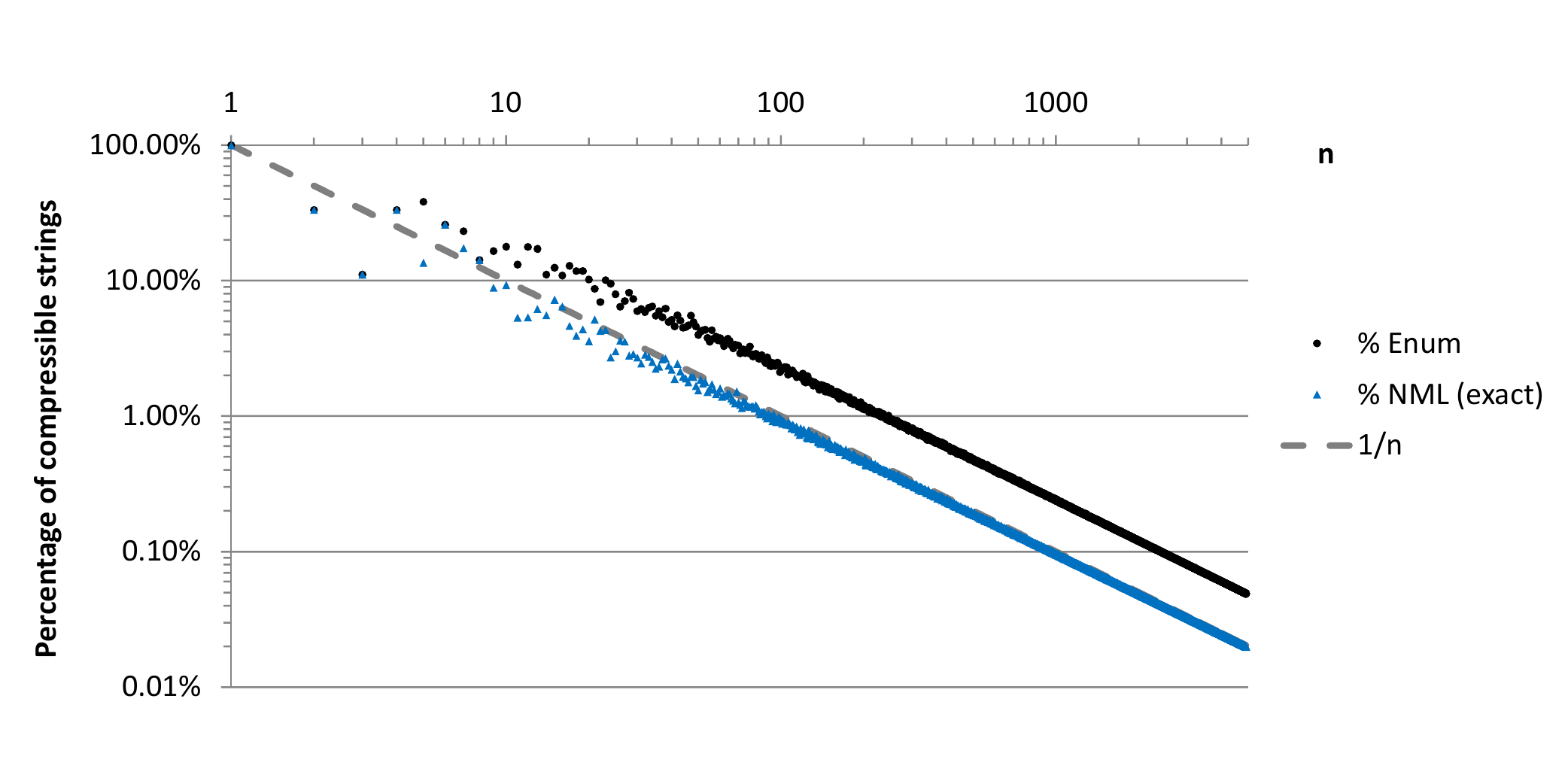}
\includegraphics[width=0.49\textwidth]{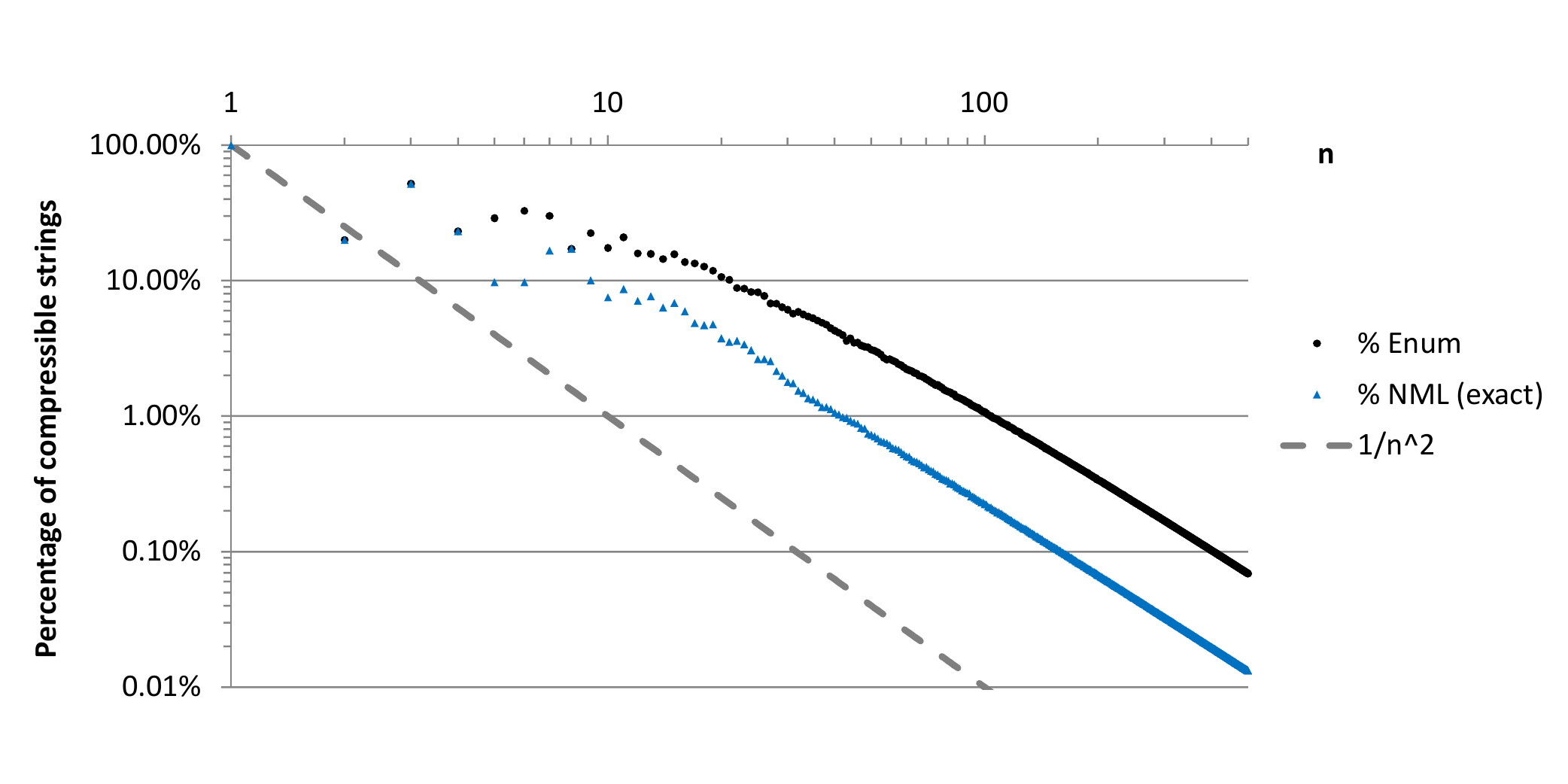}
 \caption{Percentage of compressible m-ary strings (left: $m=3$; right: $m=5$).}
 \label{fig_percentMcompressible}
\end{center}
\end{figure}

As previously, we perform an exact numerical calculation for all $n, 1 \leq n \leq n_{max}$, with $n_{max} = 5000$ for $m=3$ and $n_{max} = 500$ for $m=5$.
Figure~\ref{fig_percentMcompressible} shows that empirically, beyond the non-asymptotic case, the percentage of compressible strings decreases at a rate of $O(1/n^{(m-1)/2})$ for both codes.
However, the enumerative code always compresses more binary strings than the NML code.

\begin{figure}[!htb]
\begin{center}
 \includegraphics[width=0.39\textwidth]{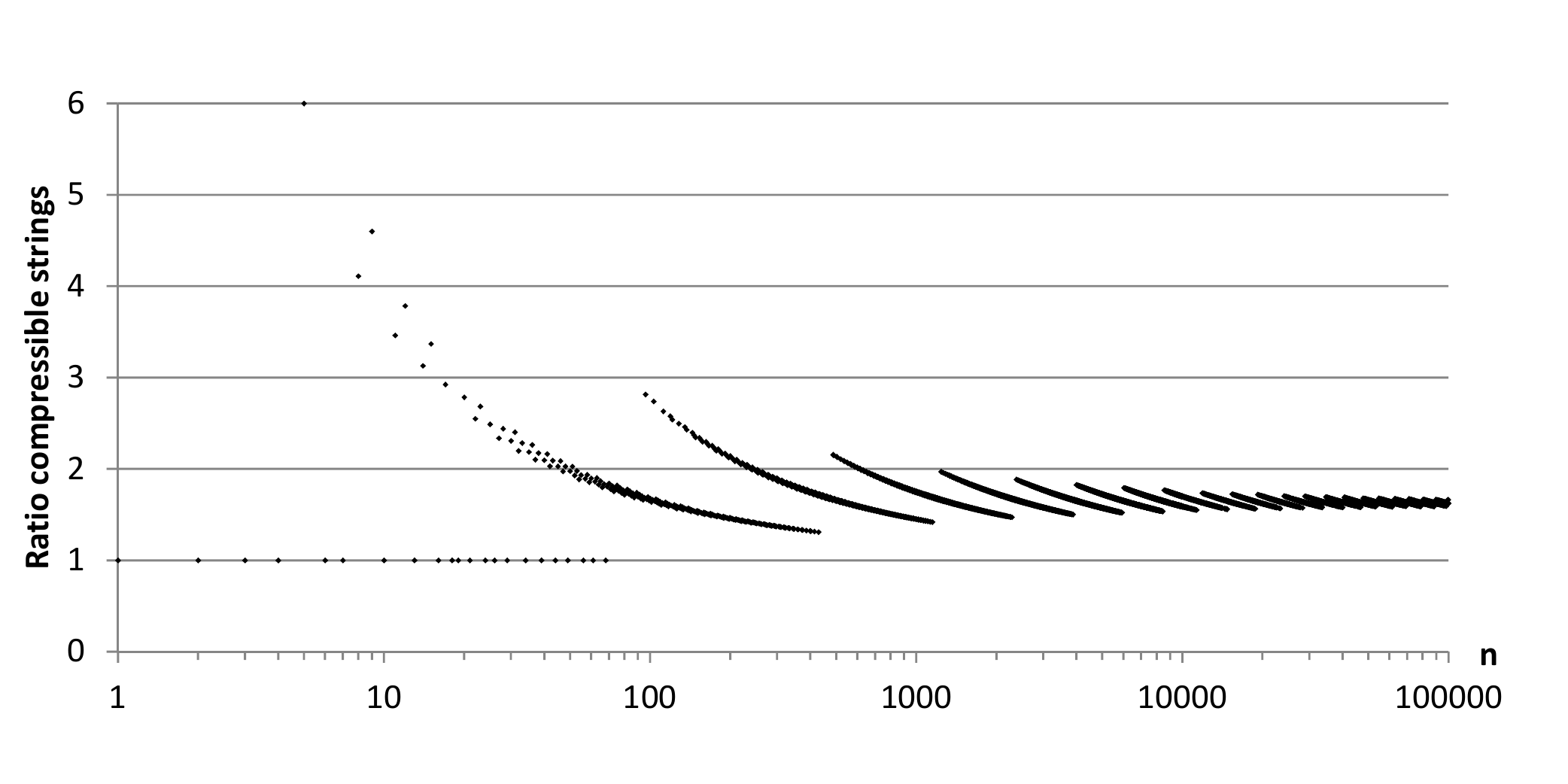}
\includegraphics[width=0.29\textwidth]{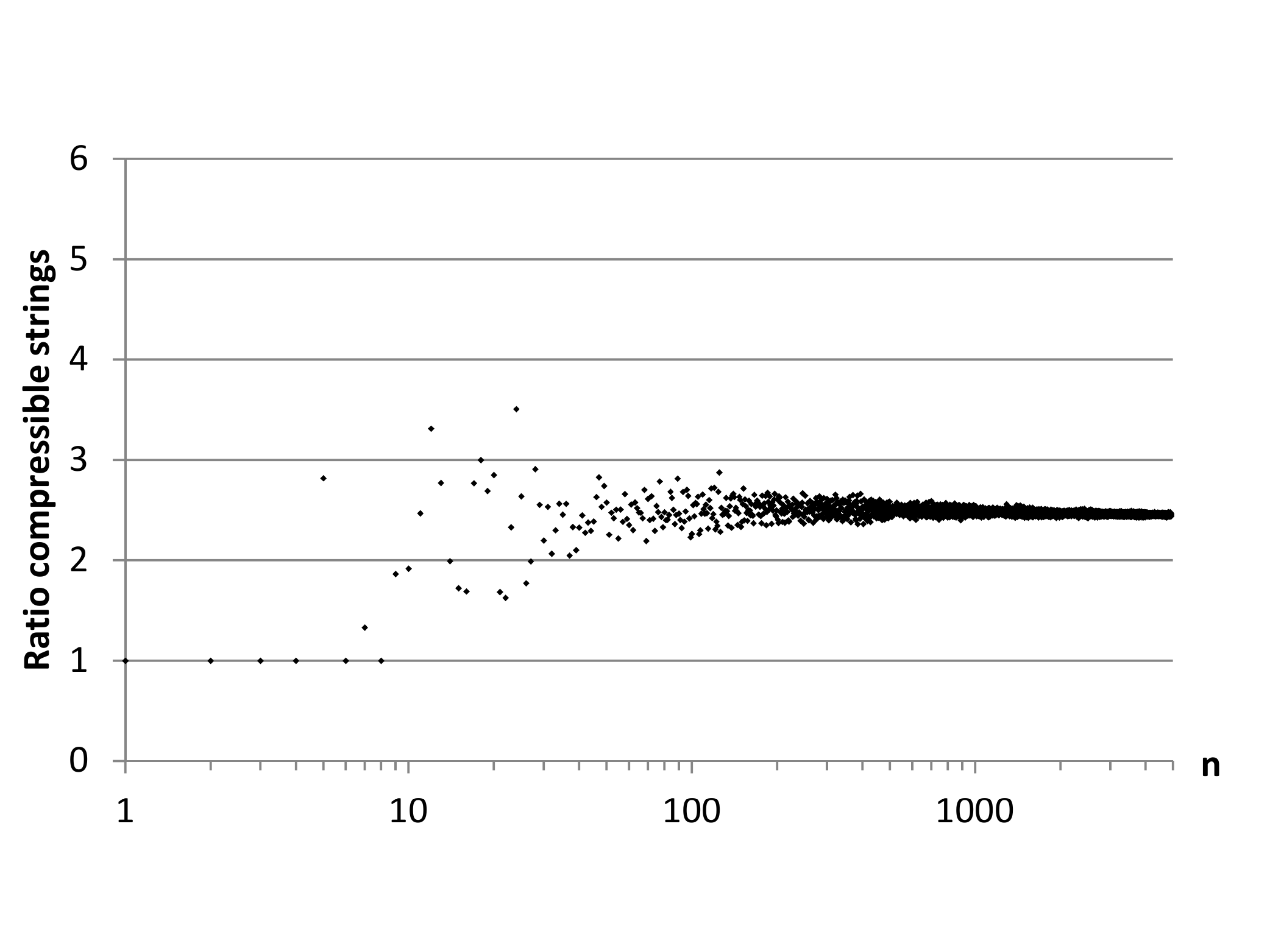}
\includegraphics[width=0.26\textwidth]{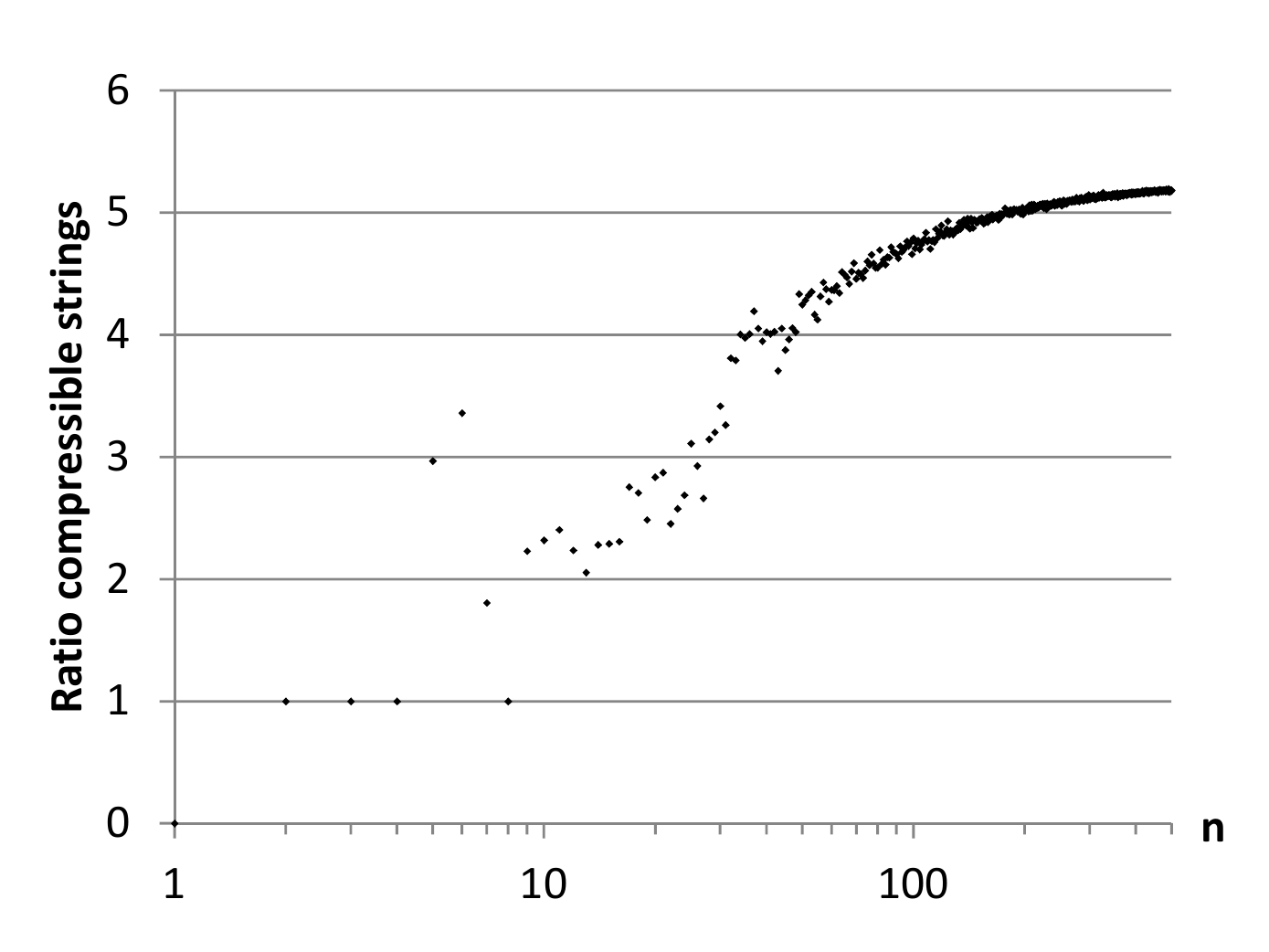}
 \caption{Ratio of compressible binary strings using the enumerative rather than the NML code (left: $m=2$; center: $m=3$; right: $m=5$).}
 \label{fig_ratiocompressible}
\end{center}
\end{figure}

To better characterize the behavior of each code, especially in the non-asymptotic case, we report in 
Figure~\ref{fig_ratiocompressible} the ratio of the number of compressible strings of the enumerative to the NML code, for $m=2$ (Bernoulli), $m=3$ and $m=5$. 
Empirically, beyond the non-asymptotic case, this ratio converges to a constant that increases with $m$: around 1.6 for $m=2$, 2.5 for $m=3$ and above 5 for $m=5$.
We expect that this empirical behavior generalizes for larger $m$, but empirical evaluation is not feasible for large $m$, even for small $n$.
On the other hand, studying the asymptotic behavior of this ratio is a non trivial task, beyond the scope of this paper.

\subsection{Detection of a biased dice}
\label{biasedDice}

We apply the previous multinomial codes to the problem of detection of a biased dice.
A fair dice is a randomizing device with $m$ outcomes that are equally likely to occur, which can be modeled using a multinomial process with equidistributed $\theta_j = \frac{1}{m}$.
Among all the possibilities of bias, we choose a simple family of peaked multinomial distributions, like those presented in Section~\ref{secOverallM}.
A peak biased dice is then determined by one single parameter $\theta_{bias} > \frac{1}{m}$, with $\theta_1 = \theta_{bias}$ and $\theta_j = \frac{1-\theta_{bias}}{m-1}, \forall j, 1 \leq j \leq m$.

The problem is to determine whether a dice is biased  given a limited sample of multinomial trials.
Given a sample $x^n$, we compute the coding length of this sample using either the NML or the enumerative code and decide that the dice is biased if its coding length is shorter than that of the random code ($n\log m$). For a given size $n$ and a code (e.g. enumerative or NML), we compute the probability of detecting the biased dice by averaging the detection over all the possible samples of size $n$.
Formally, we thus compute:

\begin{eqnarray}
\label{eqn_biasedDice}
prob^D (\theta_{bias}, n)
&=& \mathrm{E}_{M(\theta_{bias})} 
\left( \mathbb{1}_{ \left\{ L\left(\widehat{\theta}(x^n), x^n\right) < n \log m \right\} } \right) \\
&=& \sum_{\{n_1+\ldots+n_m=n\}} {\frac{n!}{n_1! n_2! \ldots n_m!} \theta_{bias}^{n_1} (\frac{1-\theta_{bias}}{m-1})^{n-n_1} 
\mathbb{1}_{ \left\{ L\left(\widehat{\theta}(x^n), x^n\right) < n \log m \right\} } }. \nonumber
\end{eqnarray}

The issue is to be able to detect a biased dice with the minimum sample size.
Using Formula~\ref{eqn_biasedDice}, we then determine the first value of $n$ where the probability of detecting the biased dice is beyond $50\%$:

\begin{eqnarray}
\label{eqn_thresholdBiasedDice}
\underline{n}^{\; D} (\theta_{bias})
&=& \min_{n \geq 10} \{prob^D (\theta_{bias}, n) \geq 50\% \}.
\end{eqnarray}

\begin{figure}[!htb]
\begin{center}
\includegraphics[width=0.49\textwidth]{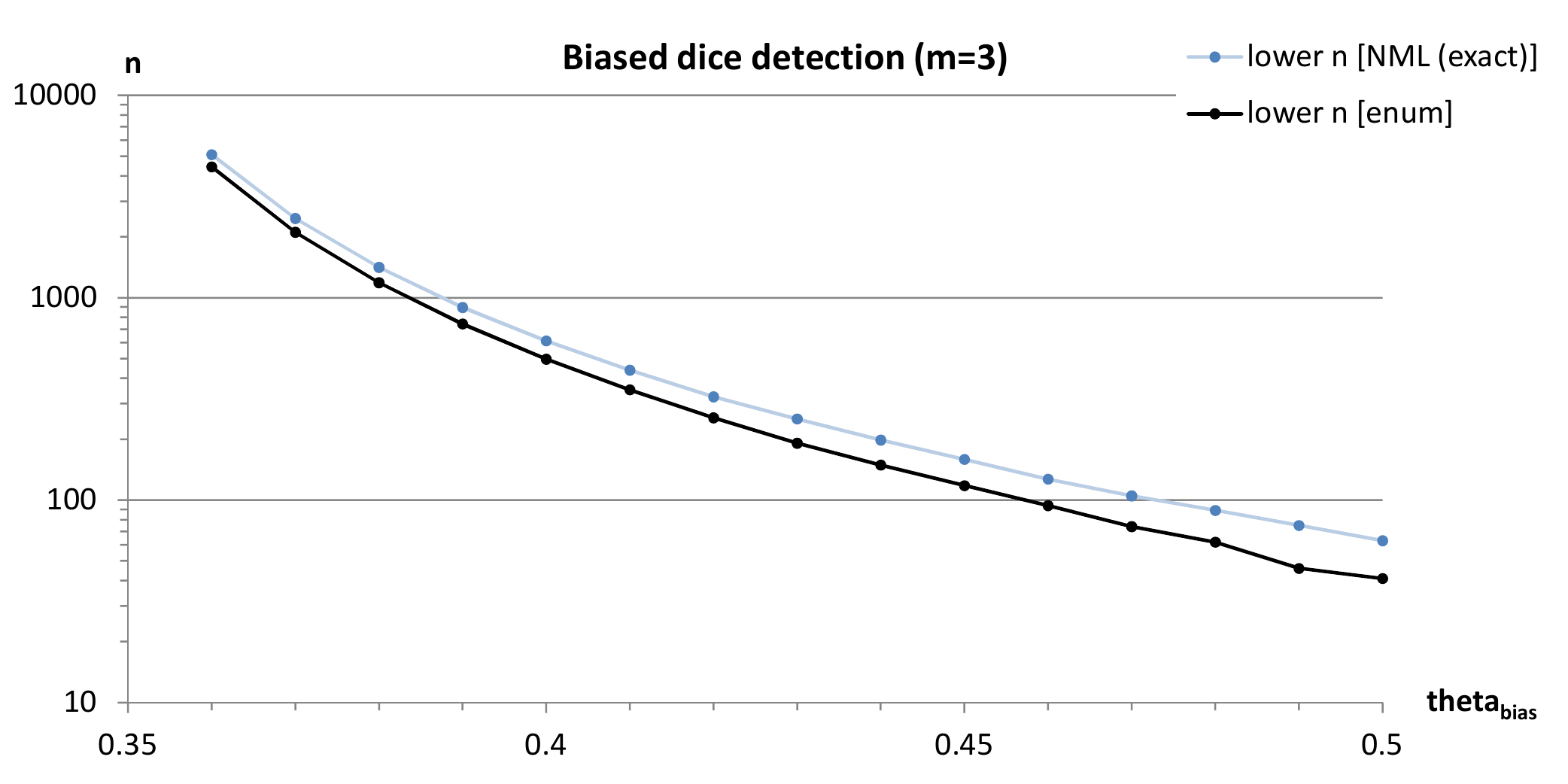}
\includegraphics[width=0.49\textwidth]{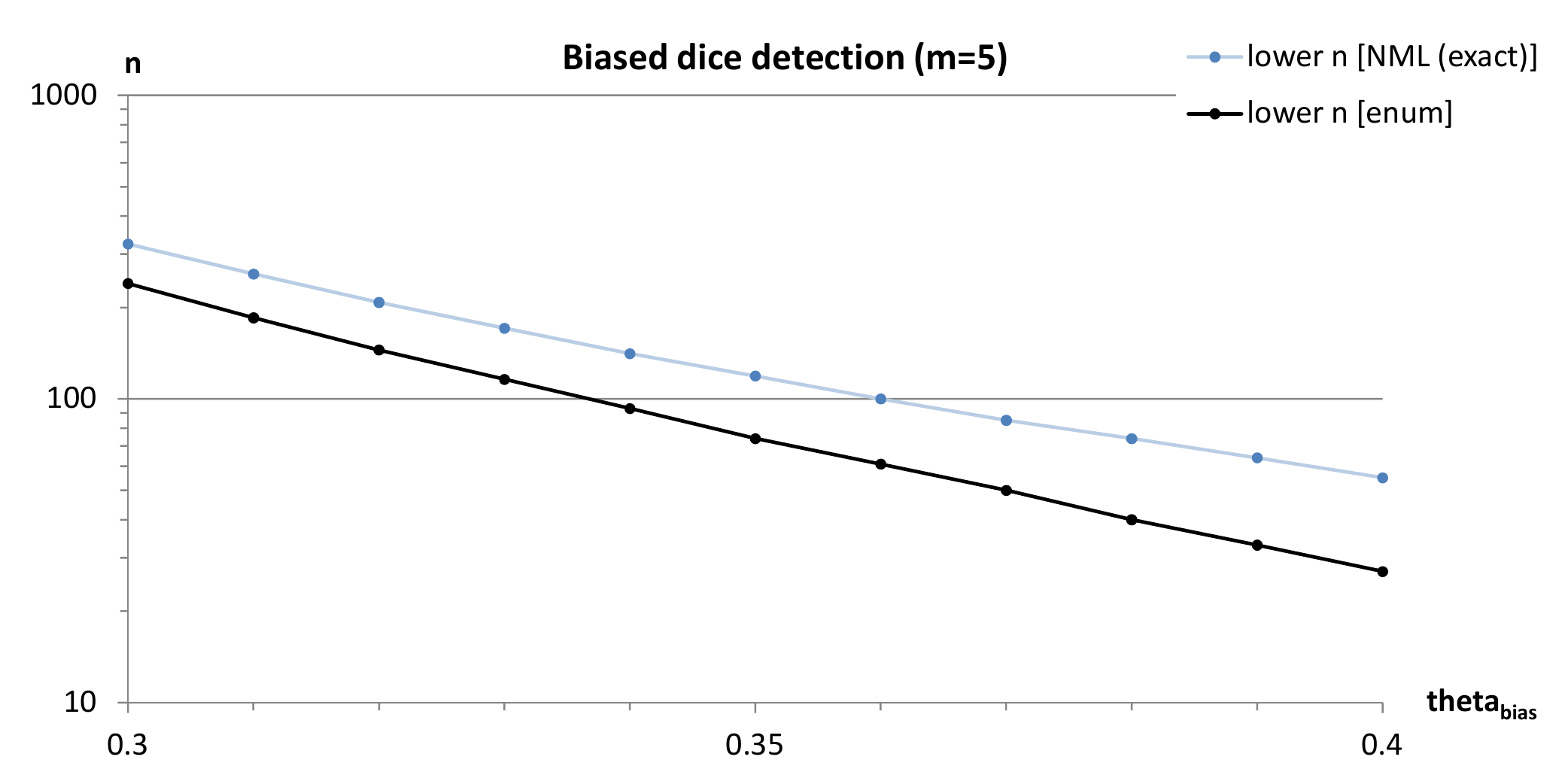}
 \caption{Minimum sample size to detect a biased dice (left: $m=3$; right: $m=5$) with probability greater than $50\%$.}
 \label{fig_thresholdBiasedDice}
\end{center}
\end{figure}

Figure~\ref{fig_thresholdBiasedDice} shows the detection thresholds computed using the NML or enumerative codes for dices with $m=3$ and $m=5$. 
As expected, the minimum number of trials necessary to detect a biased dice increases when $\theta_{bias}$ decreases.
Although all the thresholds are quite close, the enumerative code always needs smaller sample sizes to detect the biased dice. For example, for $m$=5, the enumerative code needs around $40\%$ less samples than the NML code to detect a biased dice with $\theta_{bias} \approx 0.3$.
According to the experiment, the relative difference decreases as the detection threshold increases, but this could not be studied further for heavy computational reasons. 

\subsection{Biased versus fair dice classification}
\label{secDiceClassification}

Like in the case of Bernoulli distributions, the enumerative code compresses most m-ary strings slightly better than the NML code, resulting in a better sensitivity to biased dices at the expense of more false detections in case of fair dices. 
Interestingly, the difference of behavior between the two codes increases for larger $m$.

We do not extend the coin classification experiment (see Section~\ref{secCoinClassification}) to dices, because there are multiple free parameters to define biased dice and because the calculation of the expectation of accuracy is too computationally intensive.
Still, we expect that the results might be similar, with overall a similar behavior w.r.t. the detection of biased dice, but better detection for the enumerative code in the non asymptotic case for small bias.

\section{Conclusion}
\label{secConclusion}

In this paper, we have revisited the enumerative two-part crude MDL code for the Bernoulli model, which compares favorably with the alternative standard NML code.
We have suggested a Bayesian interpretation of the enumerative code, that relies on models for finite size samples and results in a discrete definition of the likelihood of the data given the model parameter.
We have shown that the coding length of the model parameter is exactly the same as the model complexity computed by applying the NML formula using the definition of the enumerative maximum likelihood. This means that the enumerative code is both a one-part and two part code, which brings parametrization independence, optimality and simplicity.
Surprisingly, the obtained parametric complexity is twice that of the alternative classical NML code or the standard BIC regularization term. 
The enumerative code has a direct interpretation in terms of two part codes for finite sample data. The model parameter is encoded using a uniform prior w.r.t. the sample size and the data are also encoded using a uniform prior among all the binary strings of given size that can be generated using the model parameter.
This explains why the enumerative code provides a more parsimonious encoding of the data given the parameter, which compensates the larger model complexity term.
Experimental comparisons between the enumerative and NML codes show that they are very similar, with small differences only. Under the uniform distribution, the enumerative code compresses most individual sequences slightly better, resulting in a slightly better compression on average. An application to the detection of biased coins demonstrates that the enumerative code has a better sensitivity to biased coins at the expense of more false detections in case of fair coins, but the differences are small and vanish asymptotically.

Extension to the multinomial model is also presented. Using the same approach, we obtain a very simple and interpretable analytic formula for the parametric complexity term, that once again is approximately twice that of the alternative classical NML code or the standard BIC regularization term.
The resulting code, both one-part and two-part, is optimal w.r.t. NML approach and parameterization invariant, with a much simpler parametric complexity term.
It compresses most strings better than the ``standard'' NML code with a constant margin and extremely few heavily unbalanced strings with a margin logarithmic in the sample size.
Experimental comparisons extend the results obtained with Bernoulli distributions.
Both codes are very similar, with small differences that roughly increase linearly with the number of model parameters.

Altogether, the theoretical and experimental results suggest that one might use the enumerative code rather than NML in practice, for Bernoulli and multinomial distributions.

\bibliographystyle{apalike} 


\end{document}